\documentclass{ametsocV6.1}

\def\deg{${}^{\rm o}$}

\usepackage[font=small,skip=0pt,margin=0cm]{caption}
\usepackage{subcaption}
\nolinenumbers

\begin{document}

\centerline{\bf{\LARGE
A Toy Model of the Madden-Julian Oscillation
}}
\bigskip
\bigskip

\centerline{
Ian Folkins (Department of Physics and Atmospheric
Science, Dalhousie University)
}

\centerline{
Email: ian.folkins@dal.ca
}
\bigskip
\bigskip
\bigskip

%

We discuss a simple three layer model of the tropical atmosphere. The rainfall variance of the 
model is dominated by a rainfall mode moving parallel to the equator having the approximate size 
and propagation speed of the Madden-Julian Oscillation (MJO). The origin of the convective 
aggregation in the model is the imposition of distinct length scales for the deep updraft and 
stratiform downdraft circulations. Subsidence induced by the deep updraft circulation 
suppresses convective instability on a scale of $\sim$ 1000 km, while ascent induced by the 
downdraft circulation promotes convective instability on a scale of $\sim$ 500 km. Within the 
MJO envelope, high rainfall rates are maintained both by increased column relative humidity, and 
increased variance in lower tropospheric vertical motion. Each of the three model layers has a 
prescribed target pressure thickness. Convective mass fluxes introduce a mass excess into grid 
cells where there is net detrainment, and a mass deficit into grid cells from which there is net 
entrainment. Horizontal transport in the model is based on export of mass from grid cells where 
there is an excess, and import of mass toward grid cells where there is a deficit. The resulting 
patterns of horizontal convergence and divergence generate vertical motions between model 
levels. The simulated MJO events propagate eastward when there is a slight preference for mass 
deficits in the boundary layer to be compensated by inward flow from the west. The forward 
propagation of the MJO is limited by the rate at which the downdraft circulation within the MJO 
is able to generate net upward motion and promote new convective activity in advance of the 
leading edge. We also offer some guidance on how convective parameterizations that are 
implemented in models with more realistic dynamical schemes might be designed to exhibit 
stronger MJO variance.

\vfill\eject

\section{Introduction}

The Madden-Julian Oscillation (MJO) is an important mode of tropical rain variance. 
However, there does not appear to be any simple, widely accepted explanation of its 
origin, size, propagation direction, or propagation speed \citep{zhang2020}. The lack of 
such an explanation has impeded attempts to simulate the MJO in climate and weather 
forecast models \citep{hung2013,ren2021}. We present a simple explanation of the origin 
of the MJO using a three layer model of the tropical troposphere. The model is not 
directly based on the dynamical equations of motion, but is intended to provide 
conceptual guidance on how models with more realistic dynamics might be integrated with 
convective parameterizations to generate increased MJO rainfall variance.

%
%
\begin{figure}[h]
\centerline{\includegraphics[clip, trim=1.0cm 0.5cm 1.0cm 1.4cm,width=0.60\textwidth]{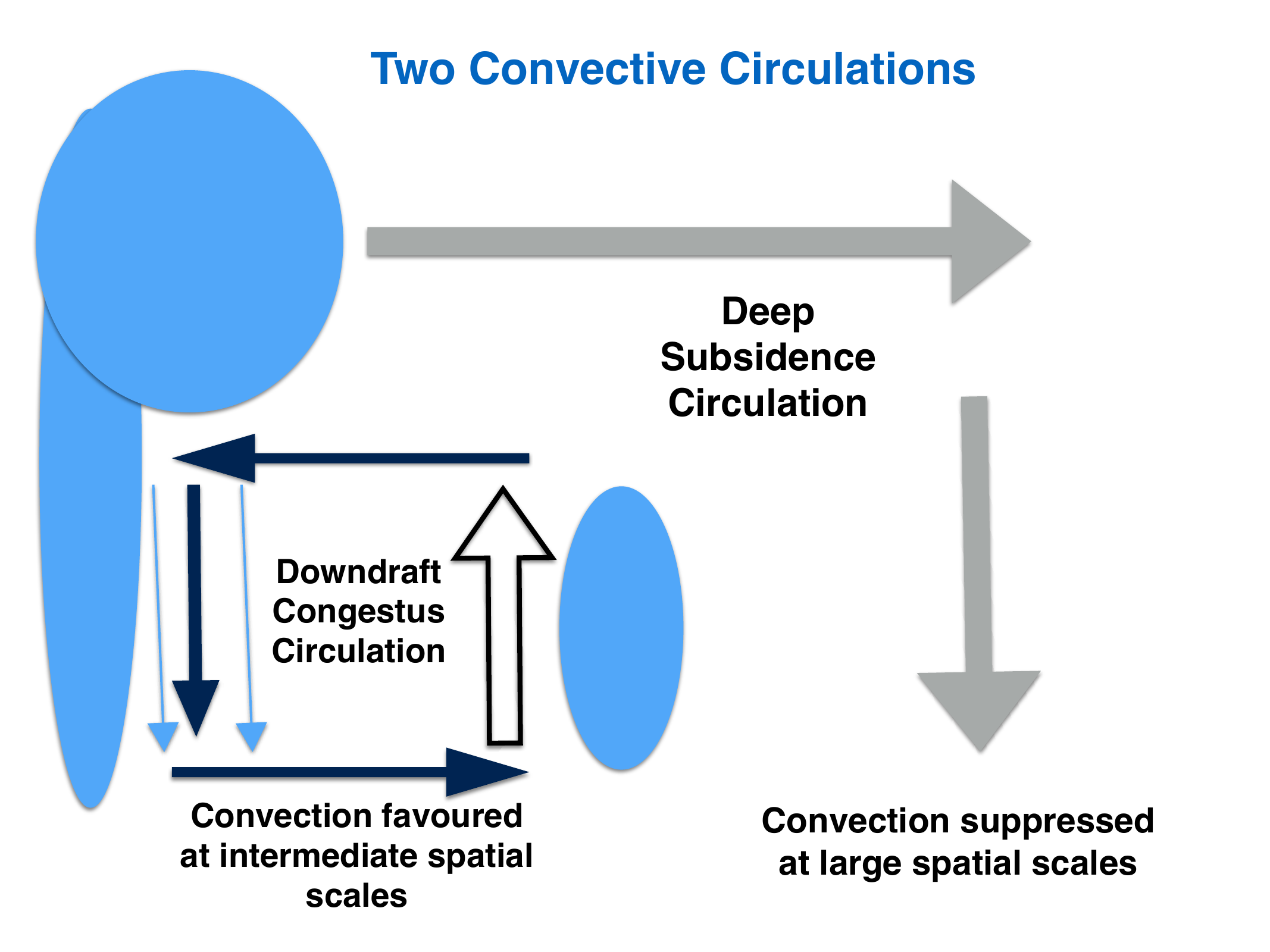}}
\caption{
Overview of the two main convective circulations in the tropics. Deep convective updrafts 
drive a large scale overturning circulation in which subsidence warming and drying inhibit 
convective development on the scale of several thousand km. Some of the precipitation 
falling from stratiform anvil clouds evaporates in the cloud free air below cloud base, 
generating downdrafts and induced ascent at intermediate spatial scales, and favoring the 
development of congestus clouds.
}\label{f1}
\end{figure}

The explanation for the origin of the MJO advanced here is illustrated in Figure 1. 
Because of the large vertical depth of deep convective heating, and proximity to the 
equator where the Coriolis parameter is near zero, the deep convective Rossby radius 
in the tropics is on the order of several thousand km. The subsidence
induced by deep convective detrainment occurs on a similar spatial 
scale, so that deep convection in the tropics gives rise to very large scale overturning 
circulations. Subsidence inhibits convective development by decreasing the relative 
humidity of the background atmosphere. Deep subsidence therefore contributes to an 
effective repulsion between deep convective cells, and a tendency to become as 
widely spaced from each another as possible. This would result in a near uniform 
rainfall distribution in the tropics, provided the overall thermodynamic sources of 
convective instability, such as sea surface temperature, and the dynamical triggers for 
convective development, were also uniformally distributed. The convective clustering 
observed within an MJO therefore requires some mechanism by which deep convection 
increases convective instability on intermediate spatial scales.

This mechanism is provided by the downdraft congestus circulation. The injection of water and 
ice condensate into the upper troposphere by deep convective clouds gives rise to stratiform 
anvils which have a much larger spatial scale than the deep convective updrafts themselves. 
Precipitation falling from these anvil clouds typically falls through unsaturated cloud free air 
below the melting level, and generates stratiform downdrafts whose mass flux can be comparable 
to the updraft convective mass flux. The vertical depth of downdraft evaporative 
cooling ($\sim$ 4 km) is much smaller than the vertical depth of deep convective heating ($\sim$ 
14 km). The spatial scale of the circulations generated by downdraft 
cooling can therefore be expected to be 
at least two times smaller than the circulations generated by
deep updraft heating. The 
induced environment uplift generated by stratiform anvil downdrafts therefore occurs at an 
intermediate spatial scale, relative to the large scale descent induced by deep 
convection. Some of this induced ascent would be expected to favor the development of cumulus 
congestus clouds \citep{johnson1999}, which, because of their smaller size, lack of internal 
organization, and short lifetimes \citep{waite2010} can be expected to be nearly in phase with 
the upward motion of the background atmosphere. To the extent that congestus clouds moisten the 
lower troposphere and serve as potential sites for the development of deep convection, they 
would act in concert with the induced downdraft uplift to favor the development of subsequent 
deep convection \citep{mapes1993}. The downdraft congestus circulation shown in Figure 1 could 
therefore provide the required mechanism for convective aggregation on the spatial scale of an 
MJO.

The explanation for the convective clustering observed in the MJO given in the previous 
paragraph is intuitively quite simple. However, the implementation of this mechanism in 
a model is not straightforward. The spatial and temporal scales of the MJO require that 
such models simulate the entire tropics. Convection resolving models of this scale 
are very computer intensive, and still require parameterizations of turbulence, radiation, 
and microphysical processes \citep{guichard2017}. Most models attempting to simulate the MJO 
have therefore relied on convective parameterizations. However, there are significant 
uncertainties in how to parameterize deep convection, cumulus convection, downdrafts, and 
stratiform anvil processes, and how to couple the convective mass fluxes generated by the 
parameterization with the grid scale variables of the model. In addition, the initial phase 
of dynamical adjustment to a heat source is mediated by waves \citep{mapes1993}. Models may 
not always resolve the waves generated by convective heat sources with sufficient accuracy 
to simulate the smaller scale downdraft congestus circulation, even if the magnitude of this 
heat source is correctly simulated by the convective parameterization. Here, we implement 
the two convective circulations MJO mechanism in a very simple three layer model of the 
tropics. The model has parameterizations for deep convection, which transports 
mass from the boundary layer to the upper troposphere, for cumulus convection, which 
transport mass from the boundary layer to the lower troposphere, and for downdrafts, 
which transport mass from the lower troposphere to the boundary layer. The effect of 
horizontal transport is to smooth out the mass anomalies generated in various grid 
cells by the vertical convective mass fluxes. In effect, it imposes by fiat the horizontal 
motions that would be expected from the wave mediated response to convective heat sources. 
For the upper troposphere and boundary layers of the model, the horizontal transport occurs 
over a spatial scale comparable to the Rossby radius of the deep convective circulation. For 
the lower tropospheric layer, it occurs over a spatial scale comparable to the Rossby 
radius of the downdraft congestus circulation. The model also has parameterizations for 
radiative descent, and for vertical motions forced by a difference in horizontal mass 
divergence between two levels.

Section 2 gives brief descriptions of the three datasets used to assess the model performance.
In Section 3, we argue that the actual horizontal length scales of the downdraft 
congestus and deep updraft circulations are smaller than would be implied by the theoretical 
expressions for their Rossby radii based on the depth of their respective heat sources. The 
fourth section is a technical description of the main model components. Although the model 
is quite simple in principle, there are some subtleties associated with the implementation 
of horizontal transport, due to the requirement that the length scales of the two convective 
circulations vary as a function of latitude. In Section 5, we discuss the main results of 
the default version of the model. These include the multiscale structure of the simulated MJO 
events, the internal circulation within an MJO which gives rise to symmetric congestus lobes 
on both sides of the equator, and the surface pressure pattern which is consistent with a 
westerly inflow toward the MJO from the trailing western edge. We also assess the realism of 
the simulated interaction between the background vertical motion and the convective mass 
fluxes, through comparison with the growth and decay of observed mass divergence profiles 
about high rain events in the Western Tropical Pacific. We then compare the spectral 
strength of the simulated MJO variance with the observed MJO variance. The eastward zonal 
speed of the simulated MJO events in the default version of the model is roughly 6 m/s. 
Somewhat surprisingly, it is difficult to find model parameters which significantly change 
this speed, without at the same time also destroying the simulated MJO variance. We derive a 
diagnostic expression for the MJO propagation speed based on the simulated rate of deep 
convective mass production just in front of the leading eastern edge of the MJO. In Section 
6, we modify various parameters to help determine which model processes are 
essential to the simulation of the MJO, and which determine the direction of zonal 
propagation. Section 7 is a discussion of some of the technical difficulties that can be 
encountered in attempting to couple the convective mass fluxes generated by the convective 
parameterization of a large scale model to the background vertical motion. Section 8 
discusses the main results and limitations of the model.

\section{Datasets}

The Tropical Rainfall Measuring Mission (TRMM) obtained rainfall estimates from five 
instruments on the TRMM satellite in combination with other satellite and rain gauge 
measurements \citep{huffman2012}. The 3B42 TRMM rainfall dataset used here has a spatial 
resolution of 0.25\deg , a time step of 3 hours, and extends from 50 \deg S to 50 \deg N.

The Integrated Global Radiosonde Archive (IGRA) is produced by the National Climatic Data 
Center (NCDC) \citep{durre2006}. We used twice daily wind data from 1998 to 2008 from six 
IGRA radiosonde stations near the equator, on the standard pressure levels of 1000, 925, 
850, 700, 500, 400, 300, 250, 200, 150, and 100 hPa.

We also used radiosonde data at five stations from the United States High Vertical 
Resolution Radiosonde Data (HVRRD) archive. The five stations used here are Koror (Palau 
Island: 7.33 \deg N, 134.48\deg E), Yap Island (9.48 \deg N, 138.08 \deg E). Truk (Moen 
Island: 7.47 \deg N, 151.85 \deg E), Ponape Island (6.97 \deg N, 158.22 \deg E), and Majuro 
(Marshall Island: 7.08 \deg N, 171.38 \deg E). The temperature anomaly of an individual 
radiosonde profile was defined with respect to monthly mean temperature profiles on a 200 m 
vertical grid.

%
%
\begin{figure}[h!]
\centerline{\includegraphics[clip, trim=0.5cm 4.1cm 0.2cm 3.2cm,width=0.80\textwidth]{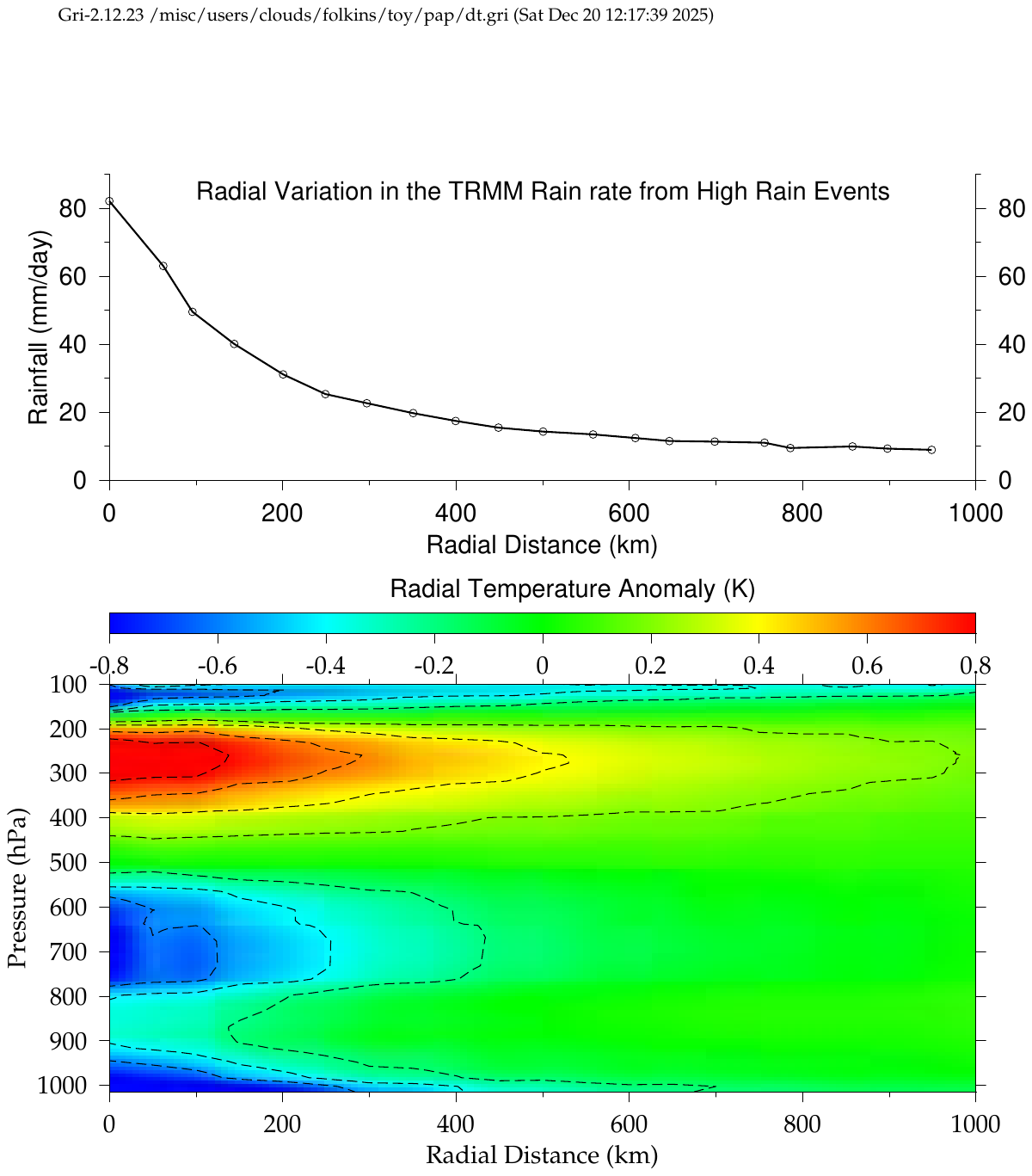}}
\caption{
The trimodal temperature response to tropical deep convection: upper tropospheric warming 
(450 hPa - 150 hPa), lower tropospheric cooling (800 hPa - 500 hPa), and boundary layer 
cooling (below 900 hPa). Deep convection also cools the Tropical Tropopause Layer (150 hPa - 
80 hPa). This temperature anomaly response pattern has motivated the choice of an upper 
tropospheric horizontal length scale $L_{UT} \sim 1100$ km, and lower tropospheric length 
scale $L_{LT} \sim 500$ km, near the equator.
}\label{f2}
\end{figure}

\section{Trimodal temperature response to deep convection}

The ratio between the updraft and downdraft mass fluxes within a deep convective system is 
likely to be strongly height dependent. However, model simulations suggest that the 
downdraft mass flux is on the order of one half of the updraft mass flux 
\citep{windmiller2023}. The convective feedback mechanism shown in Figure 1 therefore 
requires that the induced uplift from deep stratiform downdrafts be distributed over a 
substantially smaller area than the induced descent from the deep updrafts. Some indication 
of the relative spatial size of the downdraft and deep updraft circulations can be obtained 
from the radial distribution of the temperature response to strong convective events shown 
in Figure 2. This figure was adapted from an earlier publication \citep{folkins2013}, and 
was obtained by first averaging the TRMM rain rates over larger 0.5\deg \, latitude x 
0.625\deg \, longitude grid boxes. We then looked for grid boxes where the TRMM rain rate 
within any 3 hour interval exceeded 36 mm day$^{-1}$. If the radial distance of the rain 
event from one of the five Western Pacific HVRRD radiosonde stations discussed above was 
less than 1000 km, and the launch time of the radiosonde occurred within the 3 hour time 
window of the TRMM rainfall event, we calculated the radiosonde temperature anomaly profile 
with respect to the monthly mean temperature profile at that location. The temperature 
anomaly profiles of the five radiosonde stations were then combined with the relative 
distance from each rain event to construct the composite temperature response pattern shown 
in the lower plot of Figure 2. The upper plot of Figure 2 shows the the radial variation in 
the mean rainfall rate, averaged over all rain events.

Figure 2 shows that tropospheric temperature response to deep convection has three 
distinct layers: cooling near the surface (below 900 hPa), cooling in the lower 
troposphere (800 hPa - 550 hPa), and warming in the upper troposphere (450 hPa – 150 
hPa) \citep{sherwood+wahrlich1999, mapes2006, mitovski2010, virman2020}. This temperature 
response pattern results in the development of a positive stability anomaly near the 
melting level, and is associated with the preferential development of cumulus congestus 
clouds \citep{bister2004}. The use of a three layer model in this paper is partially 
motivated by the trimodal temperature response pattern shown in this figure. Higher 
rates of convective rainfall also give rise to a cooling in the tropical 
tropopause layer (150 hPa - 80 hPa). The near surface cooling can presumably be 
attributed to some combination of preferential entrainment of warmer air into deep 
convective updrafts, together with the injection of evaporatively cooled air into the 
boundary layer from the more strongly negatively buoyant downdrafts that are generated
by higher rates of convective precipitation. The existence of two distinct downdraft 
cooling maxima is consistent with model simulations showing a distinct peak in downdraft 
mass flux just below the melting level at 3.7 km, and another peak near the top of the 
boundary layer at 1.5 km \citep{windmiller2023}.

The lower panel of Figure 2 shows that the length scales of the upper tropospheric 
warming and lower tropospheric cooling are roughly 1000 km and 500 km, respectively. The 
Rossby radius of deformation $L_{RR}$ associated with a heat source of vertical depth 
$H$, in an atmosphere with a Brunt V\"{a}is\"{a}l\"{a} frequency $N$, at a latitude with 
Coriolis frequency $f$, is given by
\begin{equation}
	L_{RR} = N H/f.
\end{equation}
From the lower panel of Figure 2, the lower tropospheric cooling extends from 800 hPa 
to 550 hPa. Using the climatological variation of pressure and potential temperature 
with height at the five HVRRD radiosonde stations used to construct Figure 2, it can be 
shown that $N \sim 0.013$ $s^{-1}$ and $H \sim 3100$ m. Use of the Coriolis frequency 
appropriate for a mean latitude of roughly 7\deg \, in Eq. (1) gives $L_{RR} \sim 2000$ 
km. This theoretical estimate is considerably larger than the stratiform downdraft 
cooling circulation length scale indicated from Figure 2. This discrepancy may be
partly related to the increased relative importance of dissipation 
near the equator, where the Rossby radius would otherwise become infinite. It is also 
possible that, on the timescales comparable with downdraft formation and 
subsequent wave generation, stratiform downdraft cooling rates are more stochastic, and 
have an effectively smaller vertical extent, than the $H \sim 3100$ m implied here from 
Figure 2. Whatever the reason, we will subsequently assume that the most appropriate 
length scales for the deep updraft and downdraft circulations near the equator are those inferred from 
Figure 2, rather than the theoretical estimates based on the vertical depths of their 
heating profiles given in Eq. (1).

\section{Model description}

%
%
\begin{figure}[h]
\centerline{\includegraphics[clip, trim=1.0cm 4.2cm 1.0cm 13.5cm,width=0.90\textwidth]{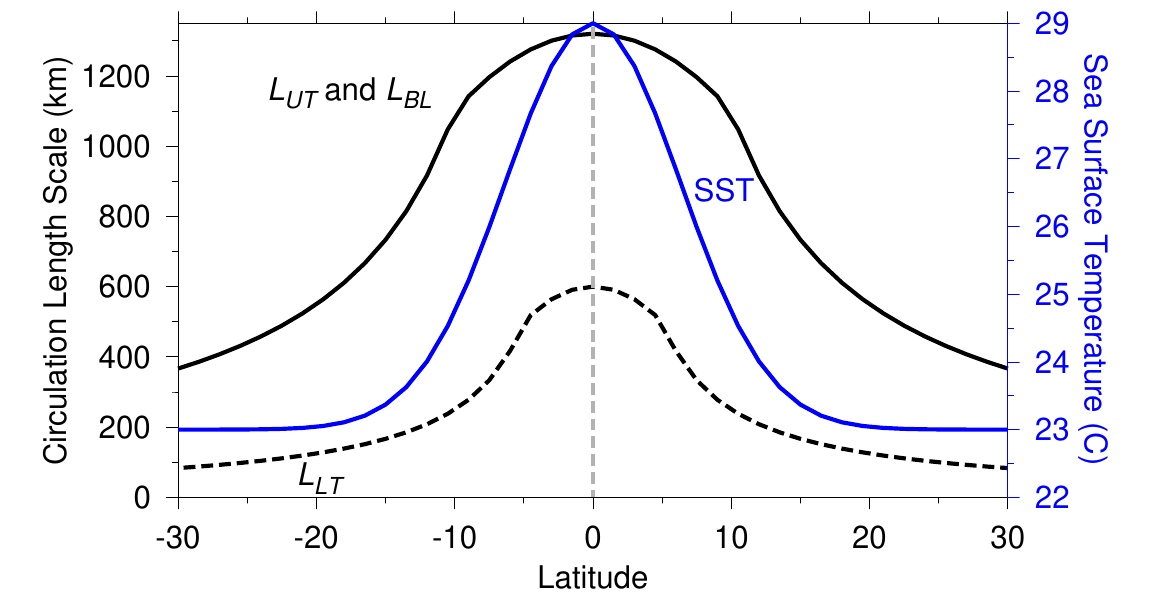}}
\caption{
The solid black curve shows the prescribed latitudinal variation of the length scale for 
horizontal transport for the upper tropospheric and boundary layers of the model ($L_{UT}$ and 
$L_{BL}$). It is therefore the spatial scale of the deep circulation. The dashed black curve 
shows the length scale for horizontal transport of the lower troposphere $L_{LT}$, and defines 
the spatial scale of the downdraft congestus circulation. The blue curve shows shows the 
latitudinal variation of the sea surface temperature.
}\label{f3}
\end{figure}

\subsection{Overall structure}

The model was restricted to within 30\deg \, of the equator. The horizontal resolution was 
1.5\deg \, in both longitude and latitude directions. The time step was 60 minutes. As shown in 
Figure 3, sea surface temperatures were fixed at 29 \deg C at the equator, decreasing with 
distance from the equator in a Gaussian manner to reach 23 \deg C at the northern and southern 
model boundaries. The three layers of the model were referred to as the boundary layer (BL), 
lower troposphere (LT), and upper troposphere (UT). The mass, or pressure difference, of the 
grid cells of a layer were permitted to vary in response to vertical or horizontal mass fluxes. 
However, grid cells in the boundary layer, lower troposphere, and upper troposphere were assigned 
target pressure difference  of 100 hPa
($\Delta p_{BL,tar} = 100$ hPa), 
400 hPa ($\Delta p_{LT,tar} = 400$ hPa), 
and 350 hPa ($\Delta p_{UT,tar} = 350$ hPa), respectively. The 
pressure at the top of the model was considered to be 150 hPa, so that the default pressure ranges 
of the boundary layer, lower troposphere, and upper troposphere were 1000 - 900 hPa, 900 hPa - 
500 hPa, and 500 hPa - 150 hPa, respectively. The actual pressure differences of the layers were
referred to as $\Delta p_{BL}$, $\Delta p_{LT}$, and $\Delta p_{UT}$. Deviations of a target grid cell from 
the target pressure difference were therefore
\begin{equation} 
\delta p_{BL} = \Delta p_{BL} - \Delta p_{BL,tar} 
\end{equation} 
\begin{equation} 
\delta p_{LT} = \Delta p_{LT} - \Delta p_{LT,tar} 
\end{equation} 
\begin{equation} 
\delta p_{UT} = \Delta p_{UT} - \Delta p_{UT,tar}. 
\end{equation} 
In general, convective mass fluxes tended to 
increase the pressure differences of grid cells from their target values. 
Conversely, horizontal mass fluxes, 
radiative subsidence, and dynamical vertical motions tended to 
relax the pressure differences of grid cells toward their target values. 
There was no background flow in the model. 

\subsection{Sigmoidal parameterization}

%
%
\begin{figure}[h]
\centerline{\includegraphics[clip, trim=0.5cm 8.3cm 0.5cm 9.5cm,width=0.80\textwidth]{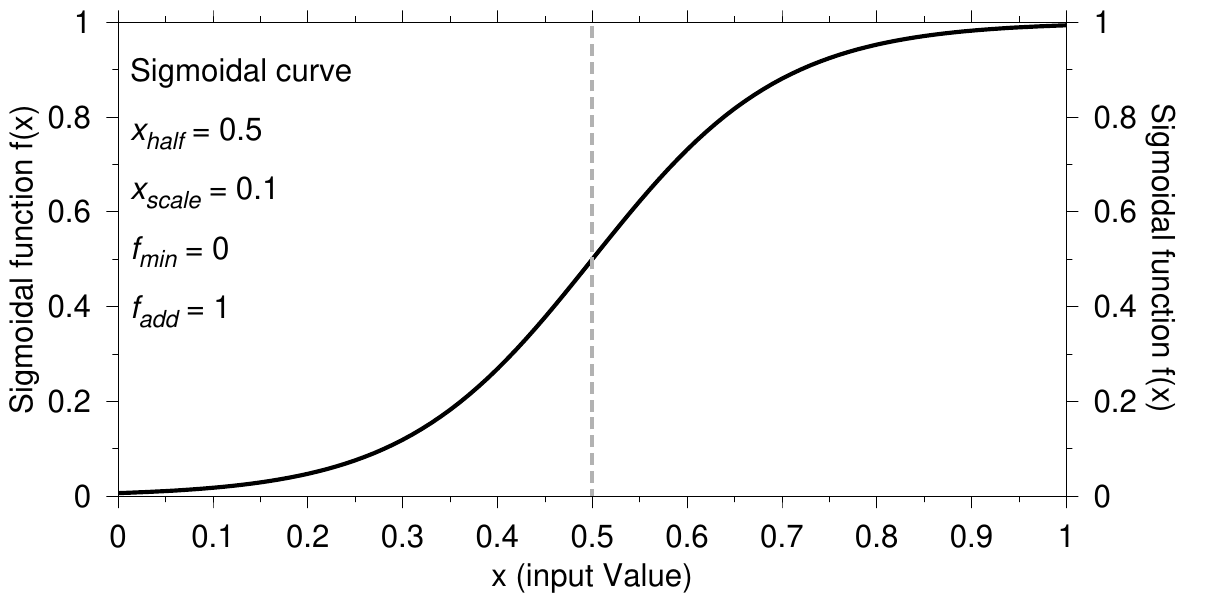}}
\caption{
The model uses sigmoidal functions to characterize the nonlinear dependence of several 
parameters on model variables. This plot shows a generic sigmoidal function $f(x)$ defined 
by four constants: the value $f_{min}$ of $f(x)$ at low values of x, the value $f_{min} + 
f_{add}$ of $f(x)$ at large values of $x$, the value of $x$ $x_{half}$ where $f(x)$ changes 
most rapidly, and a parameter $x_{scale}$ prescribing the steepness of $f(x)$ about 
$x_{half}$.
}\label{f4}
\end{figure}

Deep convective rainfall is a strongly nonlinear function of the column relative humidity 
(or column water) \citep{holloway2009}. We used sigmoidal functions to characterize this, and several
other,
nonlinear relationships in the model. These functions have the general form
\begin{equation}
f_{sig}(x) = f_{min} + {f_{add}\over{1 + e^{-x_{norm}}}}.
\end{equation}
Here, $f_{min}$ is the value of the sigmoidal function at low values of x, while $f_{min} + f_{add}$ 
is the value of the function at large values of x. $x_{norm}$ refers to the normalized value of $x$, 
\begin{equation}
x_{norm} = {{x - x_{half}}\over{x_{scale}}}.
\end{equation}
$x_{half}$ is the value of $x$ where $f_{sig}(x)$ assumes a value half way between $f_{min}$ 
and $f_{min}+f_{add}$, and is also the value of $x$ for which $f_{sig}(x)$ changes most 
rapidly. The parameter $x_{scale}$ determines the steepness of $f_{sig}(x)$ in the vicinity 
of $x_{half}$. Figure 4 shows a sigmoidal curve with particular choices for each of the four 
parameters.

\subsection{Surface heat and moisture fluxes}

The temperature and water vapor mass mixing ratio tendencies of the boundary layer due to heat 
and moisture fluxes from the ocean were defined as
\begin{equation}
dT_{BL}/dt = a_{T} (T_{SST} - T_{BL}) 
\end{equation}
\begin{equation}
dr_{v,BL}/dt = a_{r_v} (e_{s,SST} - e_{BL}).
\end{equation}
$T_{BL}$ and $T_{SST}$ refer to the temperatures of the local boundary layer and sea surface 
temperature, $r_{v,BL}$ to the local boundary layer water mass vapor mixing ratio, $e_{s,SST}$ to the 
saturated vapor pressure of the local sea surface temperature, and $e_{BL}$ to the water vapor 
pressure of the boundary layer. The parameters $a_{T} = 1 \times 10^{-10}$ $s^{-1}$ and $a_{r_v} = 
0.5 \times 10^{-6}$ ${Pa s}^{-1}$ were chosen to give reasonable values for the surface 
fluxes of sensible heat and water vapor.

\subsection{Radiative subsidence}

%
%

\begin{figure}[h]
\centerline{\includegraphics[clip, trim=0.5cm 6.0cm 0.1cm 9.7cm,width=0.80\textwidth]{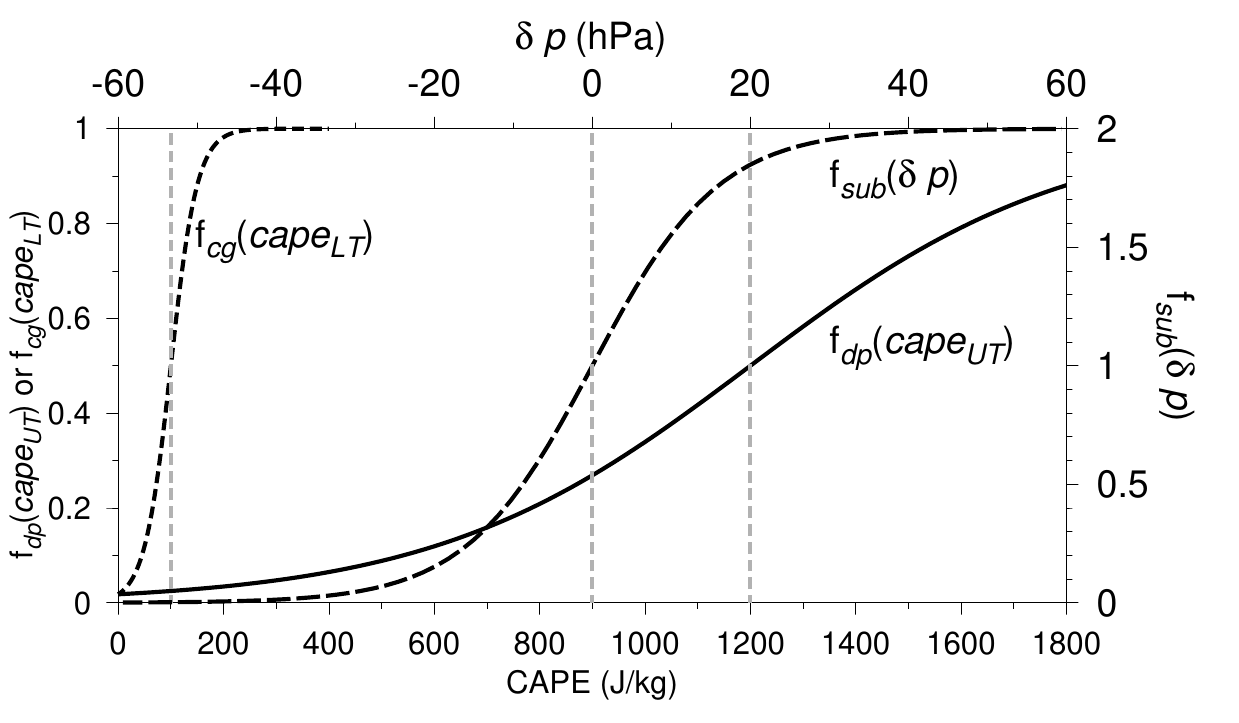}}
\caption{
Three of the sigmoidal function used in the model. $f_{cg}(cape_{LT})$ expresses the dependence 
of the congestus mass flux on lower tropospheric CAPE. $f_{dp}(cape_{UT})$ expresses the 
dependence of the deep convective mass flux on upper tropospheric CAPE. $f_{sub}(\delta p)$ 
characterizes the dependence of the radiative subsidence mass flux from the upper to lower 
troposphere on $\delta p$. This function modulates the subsidence mass flux between these two 
layers in such a way that the pressure thicknesses of grid cells do not deviate too strongly 
from their target values.
}\label{f5}
\end{figure}

Radiative descent occurs from the upper troposphere to the lower troposphere, and from the 
lower troposphere to the boundary layer. It is based on a prescribed radiative cooling rate of 
$Q_{UT} = 0.8$ K/day in the upper troposphere, and $Q_{LT} = 1.4$ K/day in the lower 
troposphere. There is also a prescribed cooling rate of $Q_{BL} = 1.5$ K/day in the boundary 
layer, but this radiative cooling does not induce radiative descent. The radiative 
subsidence mass flux from the upper to lower troposphere was given by
\begin{equation}
m_{sub,UT} = f_{sub}(\delta p) {Q_{UT} \over {g c_{pd} \sigma_{UT,LT}}},
\end{equation}
where $g$ is the rate of gravitational acceleration due to gravity, $c_{pd}$ is the specific heat 
of dry air at constant pressure, and $\sigma_{UT,LT}$ is the static stability between the upper 
and lower troposphere. 

In order to keep the pressure thicknesses of the upper and lower troposphere reasonably close to 
their target values, the radiative subsidence was adjusted using the $f_{sub}(\delta p)$ 
function. This parameter is a sigmoidally increasing function of $\delta p = \delta p_{UT} - 
\delta p_{LT}$, and is parameterized using $\delta p_{min} = 0$, $\delta p_{scale} = 8$ hPa, 
$f_{sub,min} = 0$, and $f_{sub,add} = 2$. The dependence of $f_{sub}(\delta p)$ on $\delta p$ 
with these parameters is shown in Figure 5. 
The upper tropospheric radiative subsidence is 
near zero when there is some combination of a large pressure deficit in the upper troposphere, 
or a large pressure surplus in the lower troposphere. Conversely, the radiative descent is twice 
as large when there is some combination of a large pressure surplus in the upper troposphere, or 
a large pressure deficit in the lower troposphere. The use of the $f_{sub}(\delta p)$ factor 
helped relax the pressure thicknesses of the upper and lower tropospheric grid cells toward their
target values. Radiative descent from the lower troposphere to the boundary layer 
was defined in a similar manner, using the same $f_{sub}(\delta p)$ function, and the radiative 
cooling rate and static stability appropriate for the lower troposphere.

\subsection{Length scales for horizontal transport of the two convective circulations}

As discussed earlier, the radial variation of the temperature anomaly response to deep 
convection suggests that, near the equator, the majority of deep updraft subsidence extends a 
distance of roughly 1000 km, and stratiform downdraft uplift a distance of roughly 
500 km, from strong convective events. Away from the equator, the length scale of each 
circulation would be expected to be inversely proportional to latitude. Therefore, for latitude 
values $lat(j)$ in excess of a particular $lat_{L,{UT}}$, with $j$ referring to the meridional 
index of a grid cell, the upper tropospheric horizontal transport length scale was defined as 
\begin{equation} 
L_{UT}(j) = L_{eq,UT} {{lat_{L,UT}}\over {lat(j)}}.  \quad\quad\quad\quad 
abs(lat(j)) > lat_{R,UT} 
\end{equation} 
To avoid an infinity at the equator, the upper 
tropospheric length scale within $lat_{R,UT}$ of the equator was defined as \begin{equation} 
L_{UT}(j) = L_{eq,UT} [1 + 0.2(1 - lat_{norm}^2)], \quad\quad\quad abs(lat(j)) < lat_{L,{UT}} 
\end{equation} where $lat_{norm} = lat(j)/lat_{L,{UT}}$. Figure 3 shows the latitudinal 
variation of the upper tropospheric horizontal transport length scale about the equator using 
$lat_{L,{UT}} = 10$\deg \, and $L_{eq,UT} = 1100$ km. It would also be possible to assume that 
the upper tropospheric length scale was simply constant for latitudes within $lat_{L,{UT}}$ of 
the equator. However, this assumption is potentially undesirable because it would introduce a sudden 
change in the latitude variation of the horizontal transport length scale at this latitude.

In the model, the deep convective mass flux is roughly twice as large as the congestus mass flux. 
As a result, the majority of the horizontal flow in the boundary layer is induced by the deep 
circulation. The horizontal transport length scale of the boundary layer was therefore assumed to 
be the same as that of the upper troposphere. Horizontal flow in the lower troposphere is 
mainly a response to the vertical transport of mass from congestus clouds and downdrafts. We 
therefore adopted a parameterization of the lower tropospheric length scale in which $L_{eq,LT} = 
500$ km and $lat_{L,{LT}} = 5$\deg. The variation of the lower tropospheric length scale with 
latitude is also shown in Figure 3.

\subsection{Horizontal transport}

Horizontal motion in the model was highly simplified. Instead of adopting a set of dynamical 
equations, we assumed that the effect of the horizontal flow was to export mass from grid cells 
where there was a local pressure excess, and import mass toward grid cells where there 
was a local pressure deficit. The excess mass of a grid cell $p$ was allocated to the surrounding 
$(i,j)$ grid cells using a weight function. If grid cell $(i,j)$ and  grid cell $p$ 
have the same latitude index $j$, they also have a common horizontal transport length scale 
$L_{UT}(j)$. If they are separated by a zonal distance $D_p(i,j)$, we defined a normalized 
horizontal transport distance $D_{pn}(i,j) = D_p(i,j)/L_{UT}(j)$. The weight function between 
two points having the same $j$ was then defined as a Gaussian function of this normalized distance.
\begin{equation}
W_p(UT,i,j) = exp[-{D_{pn}^2(i,j)}]
\end{equation}

Because the horizontal transport length scales are a function of latitude, 
defining the weight function between two grid cells having the same longitude
but different latitudes was more complicated. For two adjacent grid cells at the same longitude, we first 
determined their average horizontal transport length scale. We then calculated the normalized 
distance between them, and used Eq. (12) to determine their weight function. For two grid 
cells differing by more than one grid cell in the meridional direction, the weight function 
was defined as the product of all intermediary weight functions of adjacent grid cells. 
Finally, for two grid cells having differing longitude and latitude, the weight function 
between them was defined as the product of the weight function along the zonal direction 
with the weight function along the meridional direction. Figure 6 shows the spatial variation
of an upper and a lower tropospheric weight function with respect to their indicated 
start points $p$. The
decrease in the circulation length scales with distance from the equator implies that the majority
of the horizontal transport in the model occurs between 15 \deg S and 15 \deg N.

%
%
\begin{figure}[h]
\centerline{\includegraphics[clip, trim=0.1cm 8.2cm 0.1cm 9.0cm,width=0.90\textwidth]{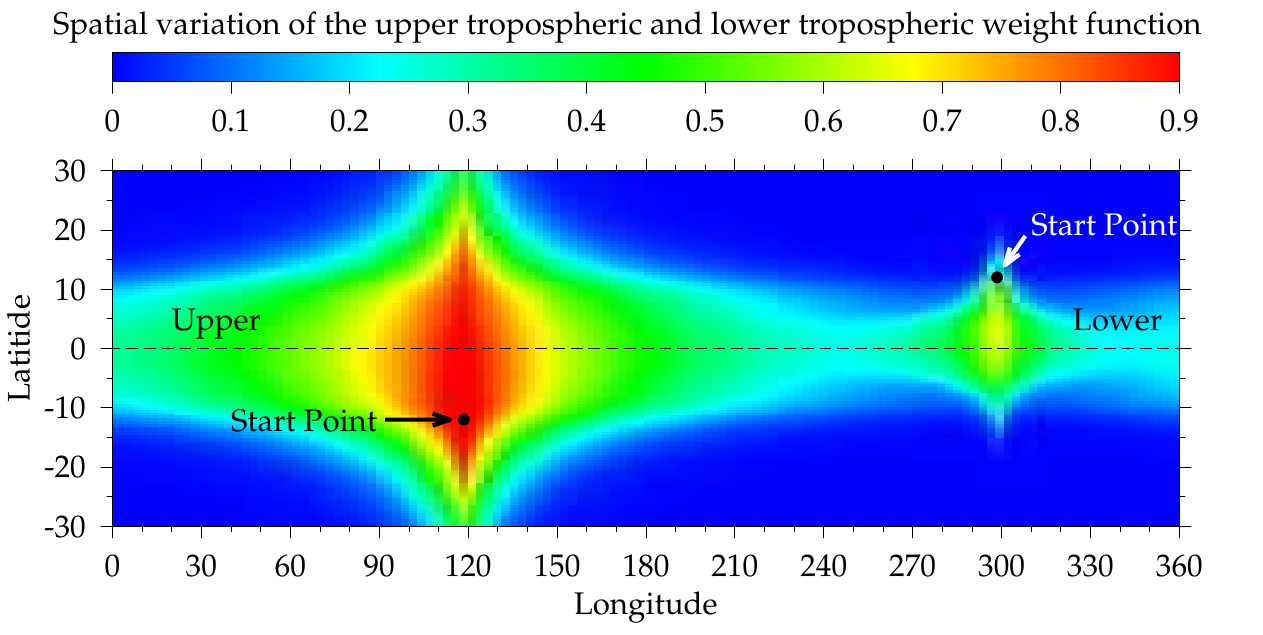}}
\caption{
Spatial variation of the upper and lower tropospheric weight functions. The weight functions 
define the spatial region over which horizontal transport occurs from a given starting point. 
The starting point for the upper tropospheric weight function was at 12 \deg S. For the lower 
tropospheric weight function, the starting point was at 12 \deg N. The length scale for 
horizontal transport increases toward the equator. This implies that the majority of horizontal 
transport toward, or away from, a grid point will be on the equatorial side of the starting 
point.
}\label{f6}
\end{figure}

Because of the earth's rotation, it is not required that the weight functions be symmetric 
in the zonal direction. Therefore, we allowed the effective horizontal transport length 
scale in the eastward direction to be slightly larger or smaller than the horizontal transport 
length scale in the westward direction. Going in an eastward direction from the original grid 
point, the length scale was multiplied by a zonal asymmetry factor $f_{L,asym}$. Going westward 
from the original grid point, the length scale was divided by this asymmetry factor.
\begin{equation}
L_{UT,east}(j) = f_{L,asym} L_{UT}(j)
\end{equation}
\begin{equation}
L_{UT,west}(j) = L_{UT}(j)/f_{L,asym}
\end{equation}
In practice, the rainfall variance of the model was extremely sensitive to small deviations of 
the zonal asymmetry factor from one. We therefore only discuss simulations in which either 
$f_{L,asym} = 1$ (no zonal asymmetry) or $f_{L,asym} = 0.95$ (slightly larger horizontal 
transport length scale in the westward direction).

To determine whether the grid cell at point $p$ should export mass to, or import mass 
from, its surrounding grid cells, we first defined the sum of all weight functions about point $p$.
\begin{equation}
W_p(UT,sum) = \sum_{i,j} W_{p}(UT,i,j)
\end{equation}
The local mean 
pressure thickness in the neighborhood of grid cell p was then defined as,
\begin{equation}
	\overline{\Delta p_{UT,p}} = {{\sum_{i,j} W_{p}(UT,i,j) \Delta p_{UT,i,j}} \over {W_p(UT,sum)} }
\end{equation}
If the pressure thickness at point $p$ deviated from this locally weighted value, the amount of 
mass exported from or imported toward point $p$ was given by,
\begin{equation}
\delta m_p = f_{move}  (\Delta p_{UT,p} - \overline{\Delta p_{UT,p}})/g,
\end{equation}
where positive $\delta m_p$ implies an export of mass, and negative $\delta m_p$ the reverse. The 
parameter $f_{move}$ determines the rapidity with which horizontal transport reduced anomalies in 
the local pressure thickness of a grid cell. For grid cells at the equator, $f_{move,eq} = 0.6$. 
For grid cells off the equator, this value was multiplied by the ratio of the integrated weight 
function about that point, relative to the value at the equator.
\begin{equation}
f_{move}  = {{ W_p(UT,sum) }\over {W_{eq}(UT,sum) }} f_{move,eq}
\end{equation}
Horizontal transport on the lower tropospheric and boundary layers was handled in an identical fashion,
except for the use of a different transport length scale in the lower troposphere.

\subsection{Vertical dynamics}

Horizontal transport generates net convergent inflow toward some grid cells, and net divergent 
outflow from other grid cells. The mass convergence $m_{conv}$ at grid cell was defined as the net 
mass per unit area added to a grid cell in one time step by horizontal transport. Upward vertical 
motion in the lower troposphere 
was assumed to be linearly proportional to the mass convergence of the 
lower level, minus the mass convergence of the upper level.
\begin{equation}
m_{dyn,LT} = f_{dyn} (m_{conv,BL} - m_{conv,LT})
\end{equation}
The dimensionless parameter $f_{dyn} = 0.24$ determines the response time of vertical transport to 
a differential rate of horizontal mass convergence between two levels. 
A similar equation was used to define the dynamical vertical motion between the lower and 
upper troposphere. With this parameterization, upward motion between levels was forced by some 
combination of convergence at the lower level and divergence at the upper level, and downward 
motion by the reverse combination. The effect of dynamical vertical motion between two levels 
was therefore to dampen changes in the relative mass of two levels caused by the horizontal
flow.

\subsection{Congestus convection}

Congestus clouds transport mass from the boundary layer to the lower troposphere. The congestus 
mass flux $m_{cg}$ was assumed to be an increasing function of the CAPE between the boundary layer 
and the lower troposphere ($cape_{LT}$) and the column relative humidity ($colrh$), and in phase 
with the total ( dynamic and radiative) vertical motion in the lower troposphere.
\begin{equation}
m_{cg} = a_{cg} f_{cg}(cape_{LT}) f_{cg}(colrh) (m_{dyn,LT} + m_{sub,LT})
\end{equation}
The parameter $a_{cg} = 3$ controls the amplitude of the congestus response to upward 
motion in the lower troposphere. The sigmoidal functions $f_{cg}(cape_{LT})$ and 
$f_{cg}(colrh)$ were used to modulate the congestus response based on available lower 
tropospheric CAPE and column relative humidity, and are shown in Figures 5 and 7, 
respectively. Congestus clouds were assumed to respond effectively instantaneously to the 
net dynamical forcing $m_{dyn,LT} + m_{sub,LT}$ (positive upward). The congestus mass flux 
was set to zero when the net vertical motion was downward.

Some of the rainfall that is generated by congestus clouds evaporates into the lower troposphere. 
The evaporated congestus rainfall fraction $f_{evap,cg}$ is a sigmoidal function of the relative 
humidity of the lower troposphere. It decreases from $f_{evap,cg} = 0.5$ at low relative humidity 
to $f_{evap,cg} = 0.0$ at high relative humidity. In addition, evaporation of congestus rainfall 
is not permitted to result in a lower tropospheric relative humidity in excess of $rh_{evap,add} = 
0.80$.

%
%
\begin{figure}[h]
\centerline{\includegraphics[clip, trim=0.5cm 6.0cm 0.5cm 10.8cm,width=0.90\textwidth]{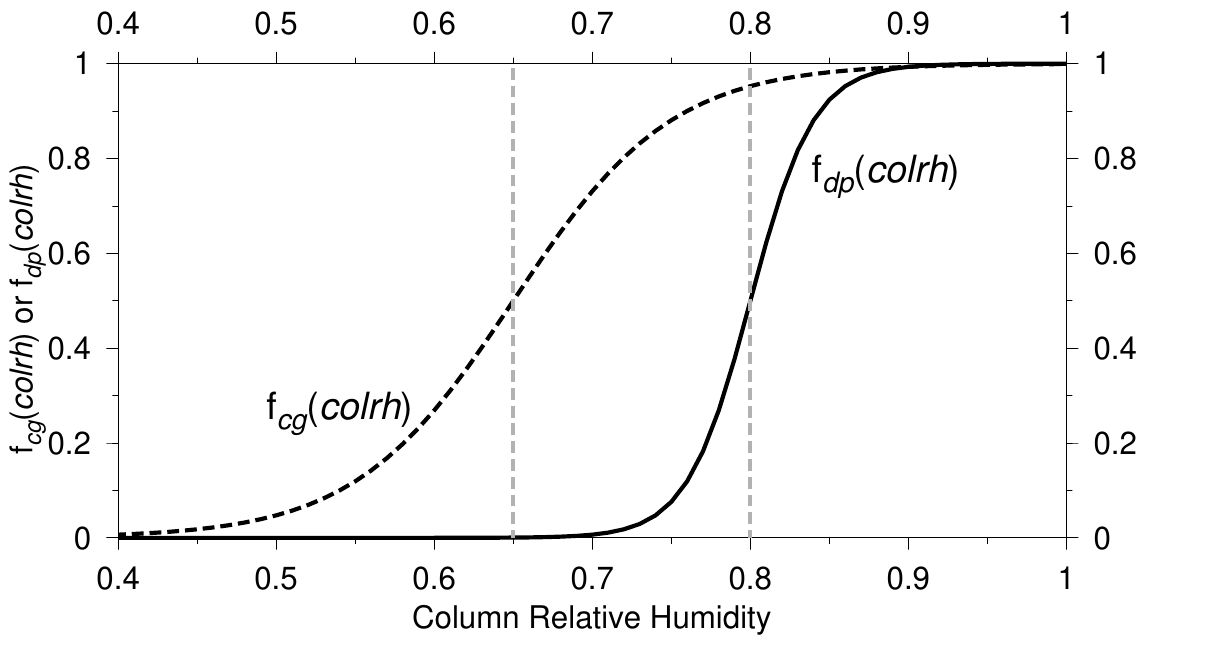}}
\caption{
Observations indicate that the congestus and deep convective mass fluxes have a nonlinear 
dependence on the column relative humidity. This is characterized in the model using sigmoidal 
functions. The nonlinear threshold column relative humidity for congestus mass was assumed to 
be lower, and the rate of increase more gradual, than for deep convective clouds.
}\label{f7}
\end{figure}

\subsection{Deep convection}

The deep convective mass flux was defined using a method very similar to the congestus mass 
flux. However, rather than responding instantaneously to the net vertical mass flux in the 
lower troposphere, we assumed that the deep convective mass flux tendency was in phase with 
the net vertical motion of the lower troposphere. Therefore, at each time step, the initial 
deep convective mass flux $m_{dp,i}$ was retained from the previous time step, and then 
adjusted by an increment proportional to ($m_{dyn,LT} + m_{sub,LT}$). For net upward motion in 
the lower troposphere, we assumed
\begin{equation}
m_{dp} = m_{dp,i} + a_{dp,+} f_{dp}(cape_{UT}) f_{dp}(colrh) (m_{dyn,LT} + m_{sub,LT}).
\end{equation}
For net downward motion in the lower troposphere, we assumed
\begin{equation}
	m_{dp} = m_{dp,i} + a_{dp,-} [1. - f_{dp}(cape_{UT}) f_{dp}(colrh)] (m_{dyn,LT} + m_{sub,LT}).
\end{equation}
We used $a_{dp,+} = 6$ for upward lower tropospheric motion, and $a_{dp,-} = 10$ for 
downward lower tropospheric motion. The functions $f_{dp}(cape_{UT})$ and $f_{dp}(colrh)$ 
were used to modify the change in $m_{dp}$ to vertical motion
depending on the local value of upper tropospheric CAPE
and the column relative humidity. These were again parameterized using 
sigmoidal functions and shown in Figures 5 and 7. Rainfall from deep convective clouds 
was also allowed to evaporate in the lower troposphere, with the parameterization being 
identical to that used for cumulus convection.

Downdrafts were generated by the evaporation of deep convective rainfall, and transported mass 
from the lower troposphere to the boundary layer. The maximum downdraft mass flux was a 
prescribed fraction $f_{dn}$ of the deep updraft mass flux. This fraction increased from 
$f_{dn} = 0$ at low values of deep mass flux, to $f_{dn} = 0.8$ at large values of deep 
updraft mass flux. The evaporation of deep rainfall used in generating this downdraft mass flux was 
assumed to be the minimum required to generate a negative downdraft buoyancy of 
$b_{dn} = -0.01$ m/s$^2$ in the boundary layer, with the total deep rainfall evaporation 
fraction not allowed to exceed $0.6$. The downdraft mass flux was set to zero if the lower 
tropospheric relative humidity was larger than $rh_{dn,add} = 0.85$.

\subsection{Drizzle and anvil rain}

The relative humidity of the boundary layer and the upper troposphere layers had a tendency to
exceed realistic values. We prescribed a maximum boundary layer relative humidity $rh_{BL,max} = 
0.85$, and a maximum upper tropospheric relative humidity $rh_{UT,max} = 0.90$. The excess 
moisture at grid cells that exceeded these limits was converted to drizzle rain and anvil rain 
respectively. This rainfall production was done in a way which conserved moist enthalpy, and 
therefore resulted in an increase in the temperature of the boundary layer or upper troposphere.

\section{Results}

%
%
\begin{figure}[h]
\centerline{\includegraphics[clip, trim=0.1cm 7.5cm 0.1cm 7.0cm,width=0.90\textwidth]{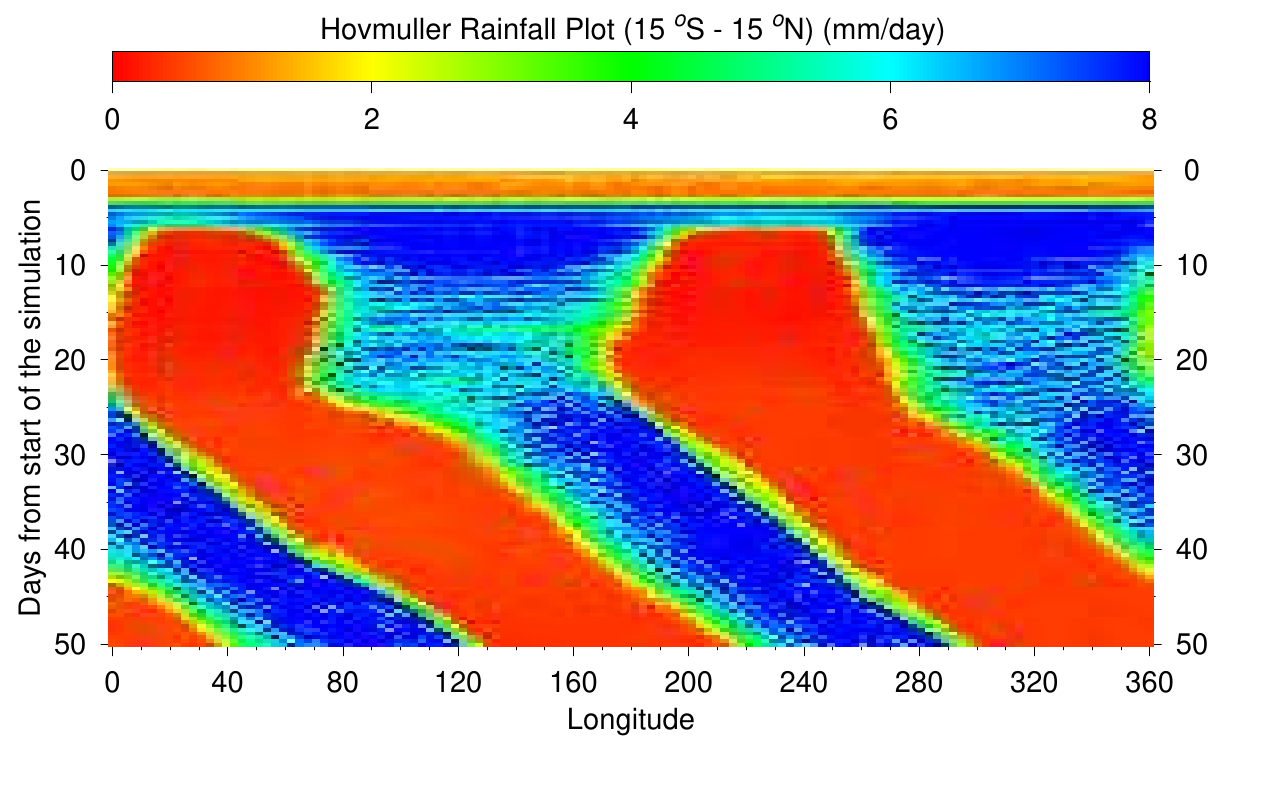}}
\caption{
Hovm\"{o}ller diagram of the latitudinally averaged (15 \deg S - 15 \deg N) total rainfall
for the first 50 days of the default version of the model. 
There is a rapid increase in rainfall rate near day 3, when the column relative
humidity becomes sufficiently large to support convection. The rainfall is initially 
zonally uniform. Large scale clusters of enhanced rainfall appear at day 5. These clusters
are initially stationary, but begin to propagate eastward at day 25.
}\label{f8}
\end{figure}

\subsection{
Hovm\"{o}ller rainfall diagram
}

Figure 8 shows a Hovm\"{o}ller diagram, in which the total rainfall rate of the model was
averaged between 15 \deg S and 15 \deg N. The zonal asymmetry parameter was assigned a value 
$f_{L,asym} = 0.95$. In this case, the induced horizontal mass flux in response to a local 
mass surplus or deficit was slightly larger in the western direction. There is an initial 
adjustment period of several days in which the rainfall rate is near zero. This occurs because 
the initial relative humidity of the model is too low to support convection. When rainfall 
first develops, it is zonally uniform. At day 5, rainfall becomes concentrated into two 
longitude bands of roughly 90\deg \, in width. Although these bands are initially stationary, 
they start to propagate eastward at day 25 with a speed of $v = 6.03$ m/s.

\subsection{
Mean MJO horizontal structure
}

%
\begin{figure}[h!]
\centerline{\includegraphics[clip, trim=0.5cm 0.5cm 0.5cm 1.3cm,width=0.85\textwidth]{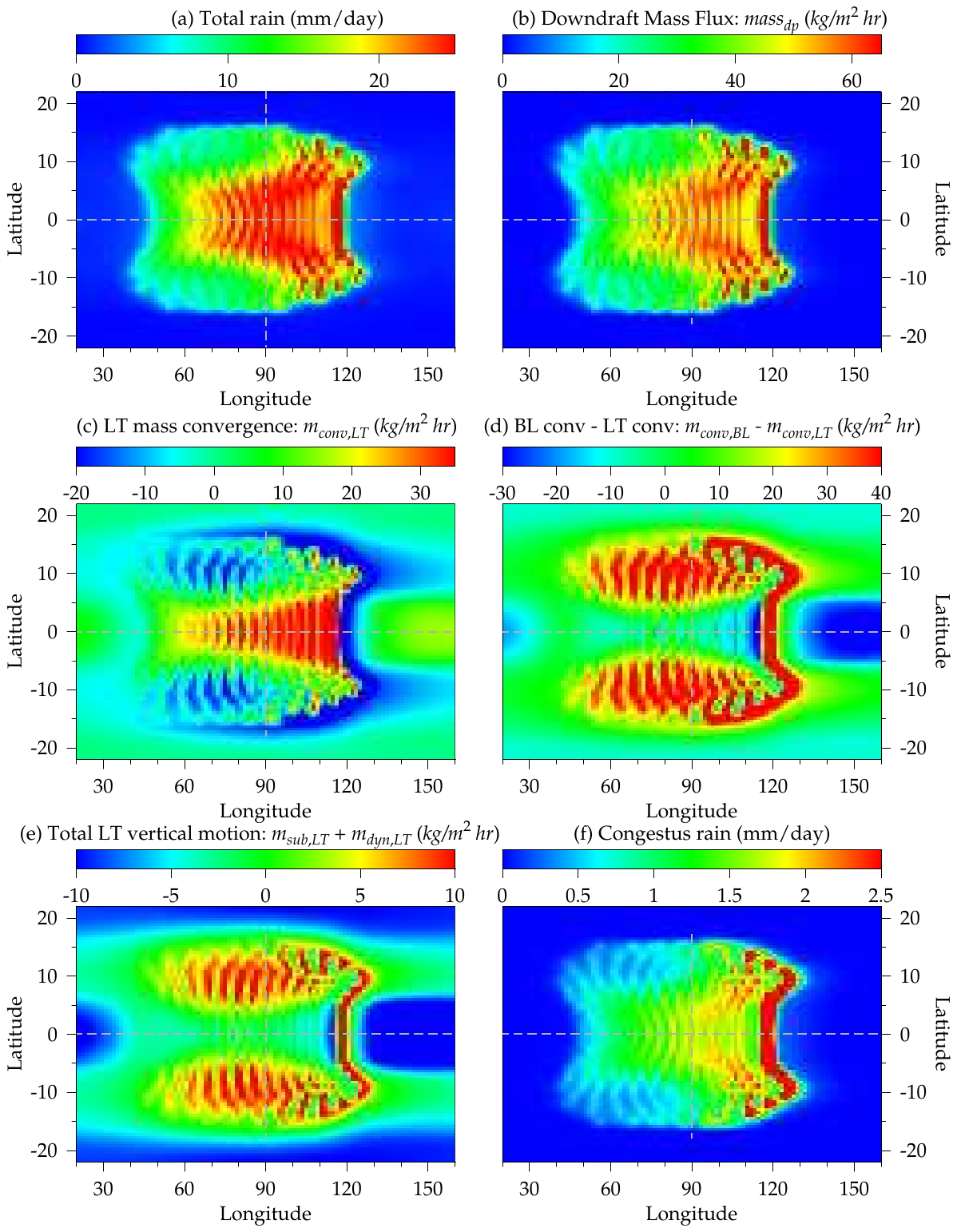}}
\caption{
Mean horizontal structure of the MJO events simulated by the model. The composites have been 
constructed by identifying, after a 25 day spinup, an MJO center at every time step of a 300 day 
model run. Model variables are then averaged about that location. Each MJO center is mapped to 
90 \deg longitude at the equator. (a) Total rain, (b) Downdraft mass flux, (c) Lower 
troposphere mass convergence. The lower tropospheric mass required by the downdraft circulation 
within the MJO is preferentially extracted from grid cells in front of the MJO. (d) Boundary 
layer convergence minus the lower tropospheric divergence. This convergence difference is 
proportional to the lower tropospheric dynamical vertical mass flux. (e) The total (subsidence + 
dynamical) vertical mass flux of the lower troposphere. (f) Congestus rainfall. This is similar 
to the total lower troposphere vertical mass flux shown in (e), but is modulated by the lower 
tropospheric CAPE and column relative humidity, which tend to be larger within the MJO.
}\label{f9}
\end{figure}

Figure 9 shows some aspects of the mean horizontal structure of the MJO events simulated by 
the model. The composite was constructed as follows. After a spinup period of 25 days, we 
identified the grid cell along the equator at each time step, between 90 E and 180 E, which 
had the largest rain rate, when averaged over all grid cells in the model within 90\deg \, to 
either side of the grid cell. This grid cell was assumed to be the center of a simulated MJO. 
We then constructed a composite MJO pattern by averaging model variables about this grid cell. 
The model simulations were run for 300 days.

The upper left panel of Figure 9 shows the total rain rate, i.e. the sum of all types of rain 
generated by the model, minus all types of evaporation. The rain rate is larger along the 
eastern side of the MJO in the direction of propagation, and larger closer to the equator. Rainfall is 
also organized into squalls lines oriented along the meridional direction. The upper right 
panel of Figure 9 shows the downdraft mass flux. Most of the rainfall in the model originates 
from deep convection, so that the downdraft mass flux pattern is very similar to the total rain 
pattern. Downdrafts remove mass from grid cells in the lower troposphere and transport this 
mass to the boundary layer. Lower tropospheric grid cells where there is a large downdraft mass 
flux will therefore typically have a mass deficit. This mass deficit generates an inward 
horizontal mass flux from the surrounding atmosphere, extending over a spatial scale equal to 
the local lower tropospheric length scale for horizontal transport. The middle left panel of 
Figure 9 shows the convergence in the lower troposphere due to this horizontal mass flux. There 
is a region of strong convergence along the equator where there is a large inward horizontal 
mass flux toward grid cells with the largest downdraft mass flux. The nearby regions shown in 
blue indicate the divergent regions which supply the mass required to sustain the downdraft 
mass flux. The middle right panel of Figure 9 shows the convergence of the boundary layer, 
minus the convergence of the lower troposphere. As shown in Eq. (19), upward dynamical motion 
between two layers occurs in the model when there is some combination of mass convergence in 
the lower level and mass divergence in the lower level. The regions shown in red in Figure 9(d) 
can therefore be taken to roughly indicate the regions within an MJO where the upward branch of 
the downdraft congestus circulation is strongest. This circulation generates upward motion 
within two symmetric lobes each between roughly 5\deg \, and 15\deg \, to either side of the 
equator, as well as along the forward leading edge of MJO. In the model, the congestus mass 
flux is in phase with the total lower tropospheric vertical mass flux, which is the sum of the 
dynamical and radiative mass flux, $m_{dyn,LT} + m_{sub,LT}$. This quantity is shown in the 
lower left panel of Figure 9. It is quite similar to the pattern shown in Figure 9(d), 
indicating that within an MJO, the lower tropospheric vertical mass flux is dominated by the 
dynamical component. The bottom right panel of Figure 9 shows the congestus rainfall generated 
by the model. The congestus rainfall rate is strongest at the eastward leading edge of MJO, 
consistent with previous suggestions \citep{kiladis2005, chen2020}. Figure 9(e) shows that the 
dynamical forcing for congestus cloud formation is also very strong within the two lobes in the 
middle and to the rear of the MJO. However, $cape_{LT}$ is reduced within these two lobes, so 
that the sigmoidal function $f_{cg}(cape_{LT})$ in Eq. (20) is too small to support congestus 
development, despite the strength of the dynamical forcing.

\subsection{
Mean longitudinal surface pressure variation
}

Figure 10(a) shows the mass convergence of the composite MJO pattern, averaged between 5\deg S and 
5\deg N for each of the three model levels, and plotted as a function of longitude. The 
irregularity of the mass convergence curves is due to the internal squall line structure of the 
MJO. The upper tropospheric mass divergence, shown here as a negative convergence, is roughly 
twice as large as the boundary layer and lower tropospheric mass convergence. The downdraft 
mass flux is therefore roughly one half as large as the deep updraft mass flux. The surface 
pressure tendency can be calculated from the residual of the three mass convergence terms.  The 
lower plot of Figure 10 shows this residual in units of hPa/day. As would be expected, this 
tendency is negative in the leading eastern half of the MJO, and positive in the rearward western 
half of the MJO. The black curve shows the surface pressure anomaly, defined in the model as a 
deviation from 1000 hPa. Due to deep convective detrainment, upper tropospheric grid cells 
within the MJO have a mass surplus. This mass surplus generates a large scale upper 
tropospheric horizontal mass export. In the leading eastern half of the MJO, this mass export 
is larger than the inward transport of mass occurring on the BL and LT levels, generating a net 
negative surface pressure tendency. In the rearward western half of the MJO, the situation is 
reversed, with the surface pressure tendency becoming positive, and generating a positive 
surface pressure anomaly to the rear of the MJO.

%
%
\begin{figure}[h]
\centerline{\includegraphics[clip, trim=0.2cm 4.9cm 0.1cm 2.3cm,width=0.90\textwidth]{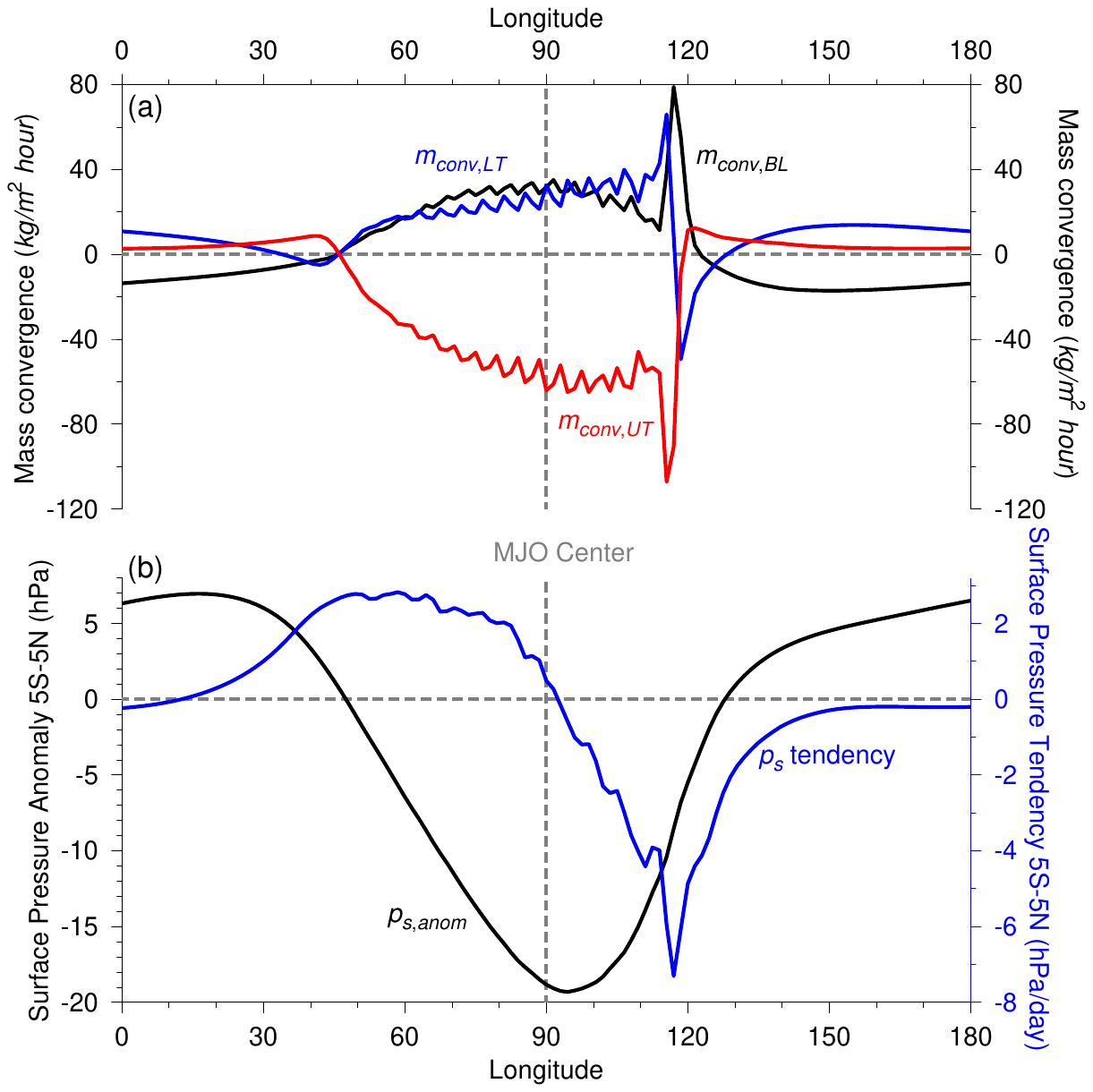}}
\caption{
(a) Longitudinal variation along the equator (5 \deg S - 5 \deg N) of the mass convergence
of each layer due to horizontal transport, obtained from the MJO composite pattern. Within
the simulated MJO events, the upper tropospheric divergence is roughly twice as large as 
the convergence of the other two layers. (b) The curve shown in blue is the surface
pressure tendency (in a frame of reference moving with the MJO)
arising from a non-zero residual of the three mass convergence terms shown in (a).
The negative surface pressure tendency in the leading half of the MJO generates the large
negative surface pressure anomaly at the center of the MJO, shown in black.
}\label{f10}
\end{figure}

The left panel of Figure 11 shows the composite MJO surface pressure anomaly pattern. There is a 
strong negative surface pressure anomaly to the east of the MJO center. Although the model does 
not have explicit horizontal winds, winds in the boundary layer can be calculated from the 
assumptions that the wind field is steady, that there is a three way force balance between the 
pressure gradient, Coriolis, and frictional accelerations, and that the radius of curvature of 
the wind field is sufficient large that the centripetal acceleration can be ignored. The 
amplitude of the zonal wind field shown on the right of Figure 11 was calculated using an 
acceleration due to friction equal to $-\epsilon {\bf V}$, with ${\bf V}$ the wind vector and 
the friction coefficient $\epsilon = 5 \times 10^{-5} s^{-1}$. As would be expected, the zonal 
pressure gradient pattern is associated with westerly winds to the rear of the MJO and easterly 
winds at the leading edge of the MJO. In the model, the relative strength of the easterly and 
westerly winds is affected by the zonal asymmetry parameter $f_{L,asym}$, with stronger westerly 
winds developing when the zonal asymmetry parameter is smaller (i.e. increased horizontal import 
of boundary layer mass into the MJO from the west).

%
%
\begin{figure}[h]
\centerline{\includegraphics[clip, trim=0.2cm 13.5cm 0.2cm 6.9cm,width=0.90\textwidth]{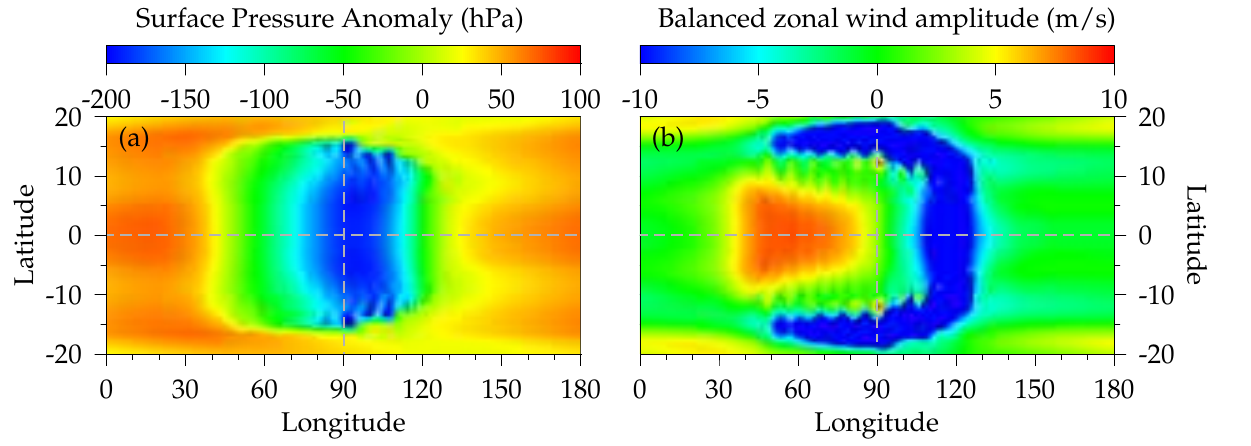}}
\caption{
(a) Spatial variation of the surface pressure anomaly about the MJO composite. There is a negative
pressure anomaly in the leading half of the MJO. (b) Amplitude of the zonal wind in the
boundary layer, assuming a three way balance between the zonal pressure gradient, Coriolis, and
frictional accelerations. There is a strong westerly inflow at the rear of the composite MJO.
}\label{f11}
\end{figure}

\subsection{Mass divergence about high rain events}

%
%
\begin{figure}[h!]
\centerline{\includegraphics[clip, trim=0.5cm 14.2cm 0.0cm 0.0cm,width=0.85\textwidth]{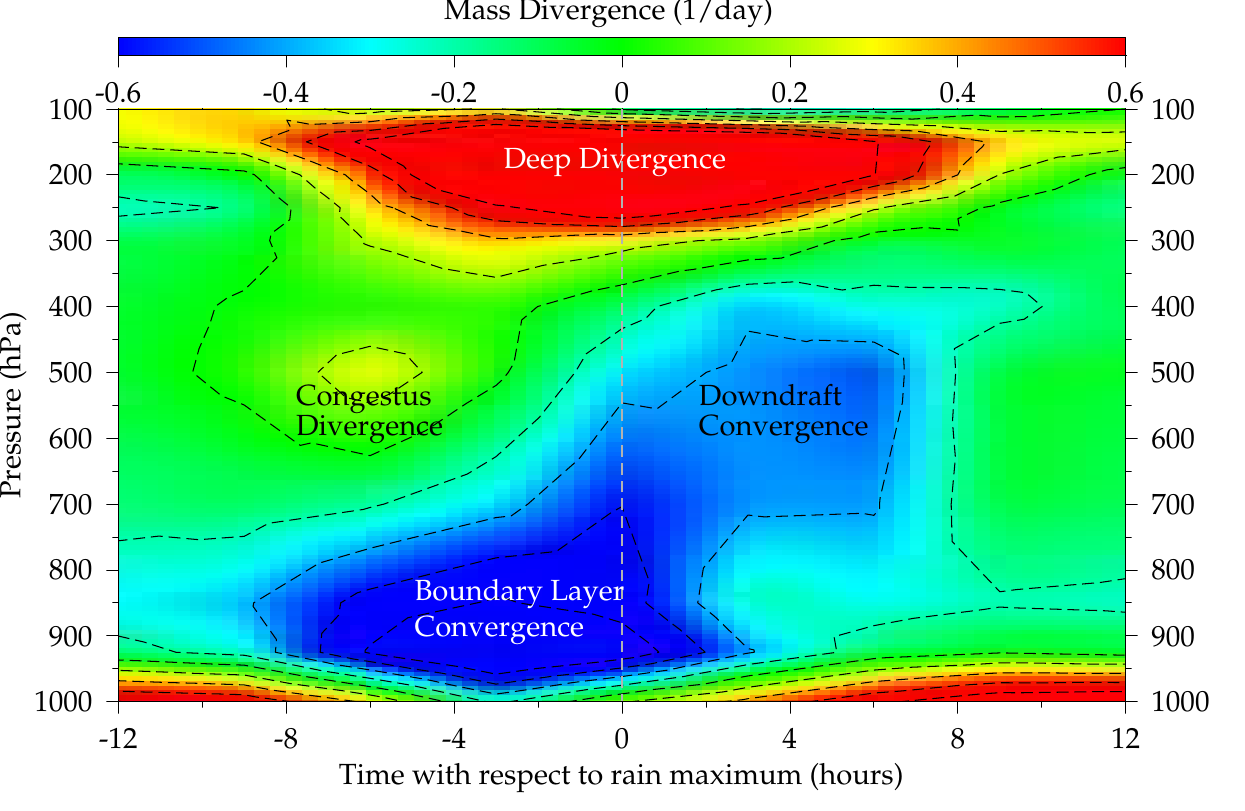}}
\caption{
Composite mass divergence pattern during the growth and decay of high rain events. It was 
obtained by combining horizontal wind profiles from triangular rawinsonde arrays in Borneo and 
the Malay Peninsula between 1998 and 2008 \citep{mitovski2010}. The mean temporal variation of 
the TRMM rain rate within the two arrays is shown in Figure 13. The mid-level mass divergence 
dipole is consistent with the tilted $\omega$ structures that usually characterize convective 
development \citep{inoue2020}.
}\label{f12}
\end{figure}

It is difficult to directly measure clear sky vertical mass fluxes in the atmosphere. However, 
mass divergence profiles about strong convective events provide some insight into the 
relationship between vertical motion and convective mass fluxes in the tropics 
\citep{mapes1995}. The mass divergence profiles shown in Figure 12 were generated by combining 
wind data from rawinsonde arrays in Borneo and 
the Malay Peninsula \citep{mitovski2010}. The spatial scale of each array ($\sim 250$ km) is 
smaller than the length scale of the downdraft congestus circulation. At any given time, the 
divergence profile of an array should therefore reflect the preferred cloud type during that 
particular phase of the downdraft circulation. 

Rain events were considered to occur at times 
between 1998 and 2008 in which the mean TRMM rain rate within each array was in the top 5 \% 
for that month. The mean TRMM rain rate profile within the two arrays is shown in Figure 13.
Simultaneous wind profiles from an array that occurred within 24 hours of the 
TRMM rain event times were then used to construct a composite diagram of the mean vertical variation
of mass divergence during the growth and decay of high rain events. The peak in boundary 
layer convergence occurs roughly 3 hours prior to peak rainfall, while the peak in upper 
tropospheric divergence occurs 1-2 hours after peak rainfall. Although these divergence 
features have a maximum amplitude of $\sim 1.4$ day$^{-1}$, we have used a smaller scale in 
Figure 12 to emphasize the mid-level divergence features. Of particular interest is the 
mid-level divergence dipole, consisting of a 500 hPa divergence peak 6 hours prior to peak 
rainfall, and a broader convergence peak centered 4 hours after peak rainfall.

%
%
\begin{figure}[h]
\centerline{\includegraphics[clip, trim=0.2cm 4.0cm 0.1cm 2.0cm,width=0.85\textwidth]{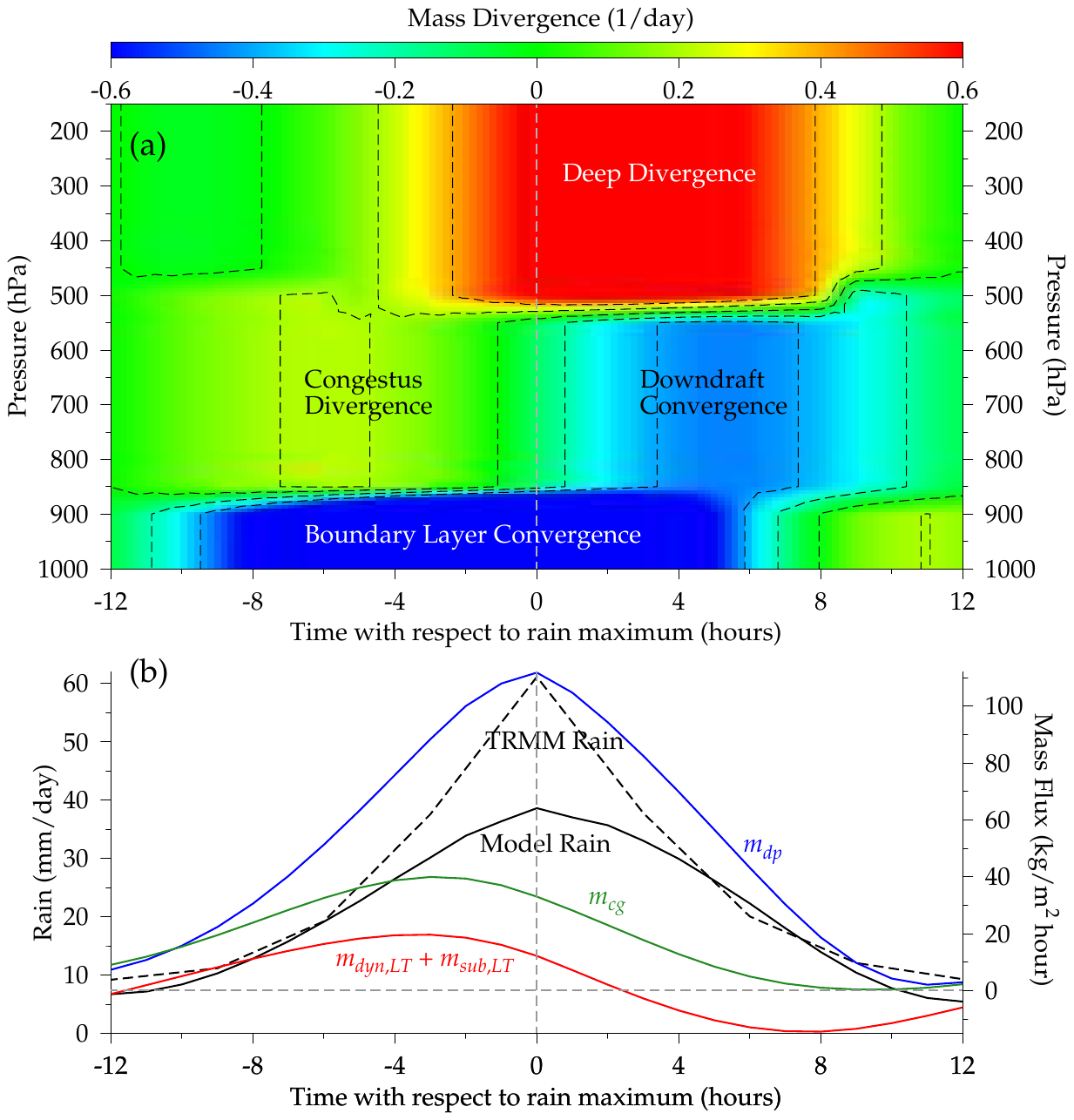}}
\caption{
(a) The top panel shows the variation in mass divergence during the growth and decay of simulated
rain events, obtained from the three model levels. (b) The dashed black curve shows the
mean TRMM rain rate variation within the two rawinsonde arrays used to construct the
observed mass divergence pattern shown in Figure 12. The solid black curve shows the
mean rain rate variation during the rain events of the model. The blue and green curves
show the variation in the deep and congestus convective mass fluxes, respectively. The
congestus mass flux, and the deep mass flux tendency, are proportional to the total 
(subsidence + dynamical) lower tropospheric vertical motion, shown in red.
}\label{f13}
\end{figure}

The mass divergence pattern shown in Figure 12 was used an observational target to guide 
the development of the convective parameterization used in the model. We considered rain 
events to occur when the total rain rate at any time step exceeded 25 mm/day. Most of these 
rain events occur within the meridional squall lines shown in Figure 9(a). We then 
used the convergent mass fluxes at the three levels, generated by the horizontal flow, 
normalized by the mass of each grid cell, to generate the divergence pattern shown in upper 
panel of Figure 13. The simulated divergence pattern is broadly similar to the observed pattern 
shown earlier in Figure 12, and in particular, appears to be in better agreement with 
observations than some larger scale models \citep{mitovski2010, mitovski2012}. The deep 
divergence of the model is somewhat lagged relative to the observations, and because of the 
limited vertical resolution of the model, the simulated mid-level congestus divergence peak is 
elongated in the vertical relative to the observed congestus peak.

The lower plot of Figure 13 shows the temporal variation of simulated total rainfall during 
the growth and decay of high rain events, together with the TRMM rain rate averaged over the 
two rawinsonde arrays. Although the two curves are similar, the TRMM rain rate is more sharply 
peaked at $t = 0$, reflecting the reduced rainfall variance of the model with respect to TRMM. 
This plot also shows the deep mass flux $m_{dp}$, the congestus mass flux $m_{cg}$, and the 
sum of the dynamic and subsidence lower tropospheric mass flux $m_{dyn,LT} + m_{sub,LT}$. By 
construction, the congestus mass flux and the deep mass flux growth rate are in phase with the 
total upward lower tropospheric mass flux. 

In the model, congestus detrainment generates 
positive mass anomalies in lower tropospheric grid cells. The parameterization for horizontal 
transport then exports this excess mass over a spatial scale equal to the local length scale 
of the downdraft circulation. This mass export generates the mid-level divergence feature 
prior to peak rainfall in the model. This simulated congestus divergence maximum occurs several hours 
prior to the peak in congestus mass flux because the downdraft mass flux (not shown) is in 
phase with the deep mass flux. Downdrafts remove mass from the lower tropospheric layer, and 
therefore progressively erode the positive lower tropospheric mass anomaly generated by 
congestus detrainment, as peak rain is approached. After peak rain, the downdraft mass flux is 
significantly larger than the congestus mass flux, and generates a progressively larger 
negative mass anomaly in the lower troposphere. This negative mass anomaly drives horizontal 
inflow from the surrounding atmosphere, which gives rise to the mid-level downdraft 
convergence feature after peak rainfall shown in Figure 13. The downdraft convergence feature 
then progressively weakens as the rain rate decays, the downdraft mass flux also dissipates, 
and the negative lower tropospheric mass anomaly generated by the downdrafts is eroded by 
inflow from the horizontal circulation.

\subsection{Rain power spectrum}

%
%
\begin{figure}[h]
\centerline{\includegraphics[clip, trim=1.2cm 5.2cm 1.5cm 2.5cm,width=0.85\textwidth]{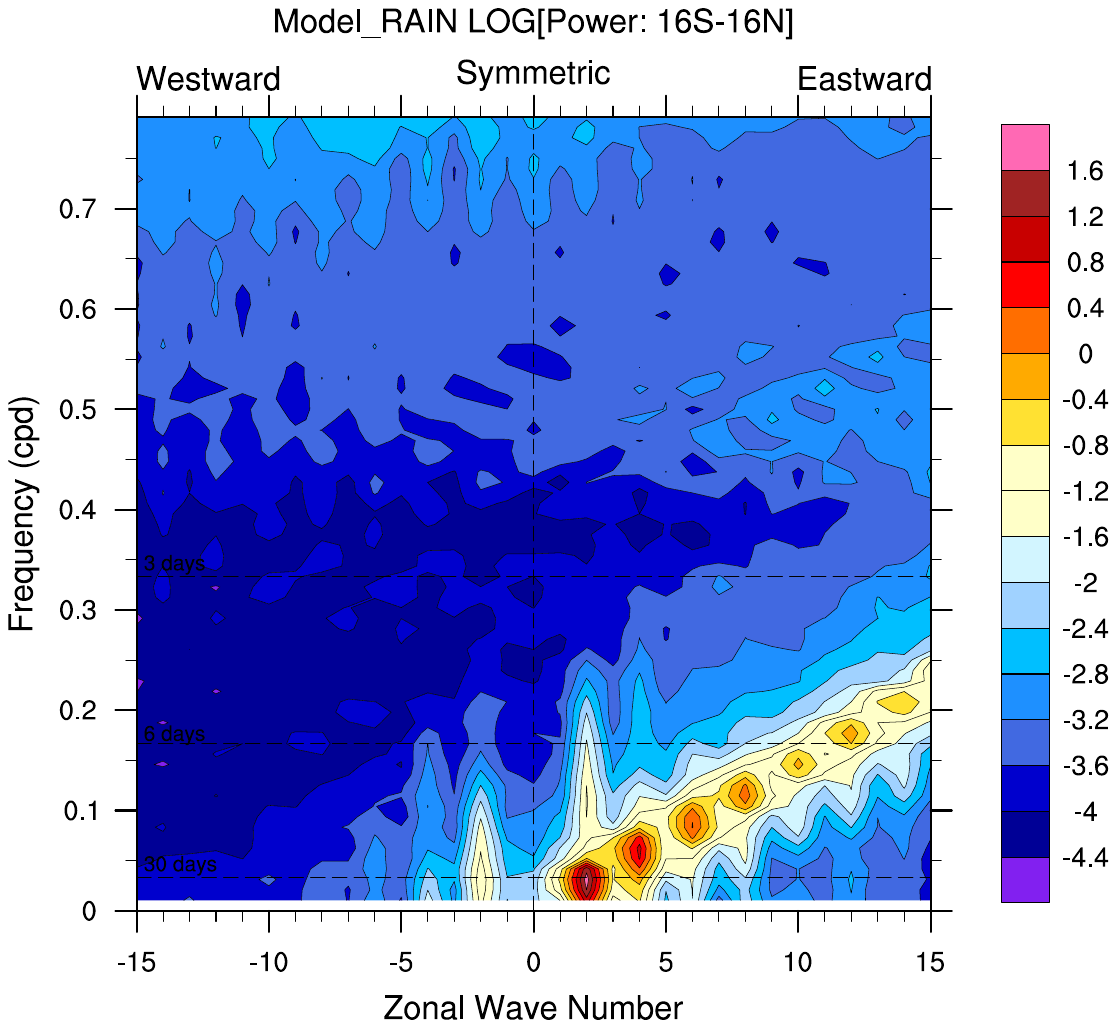}}
\caption{
Zonal wavenumber-frequency 
symmetric power spectra from the model using 300 days (minus 25 day spinup) of
16 \deg S - 16 \deg N total rainfall. Simulated MJO variance is largest for 
zonal wavenumber 2, but there are weaker harmonics at zonal wavenumbers 4, 6, 8, 10,
etc, distributed along a linear dispersion curve.
}\label{f14}
\end{figure}

%
%
\begin{figure}[h]
\centerline{\includegraphics[clip, trim=1.2cm 5.2cm 1.5cm 2.5cm,width=0.85\textwidth]{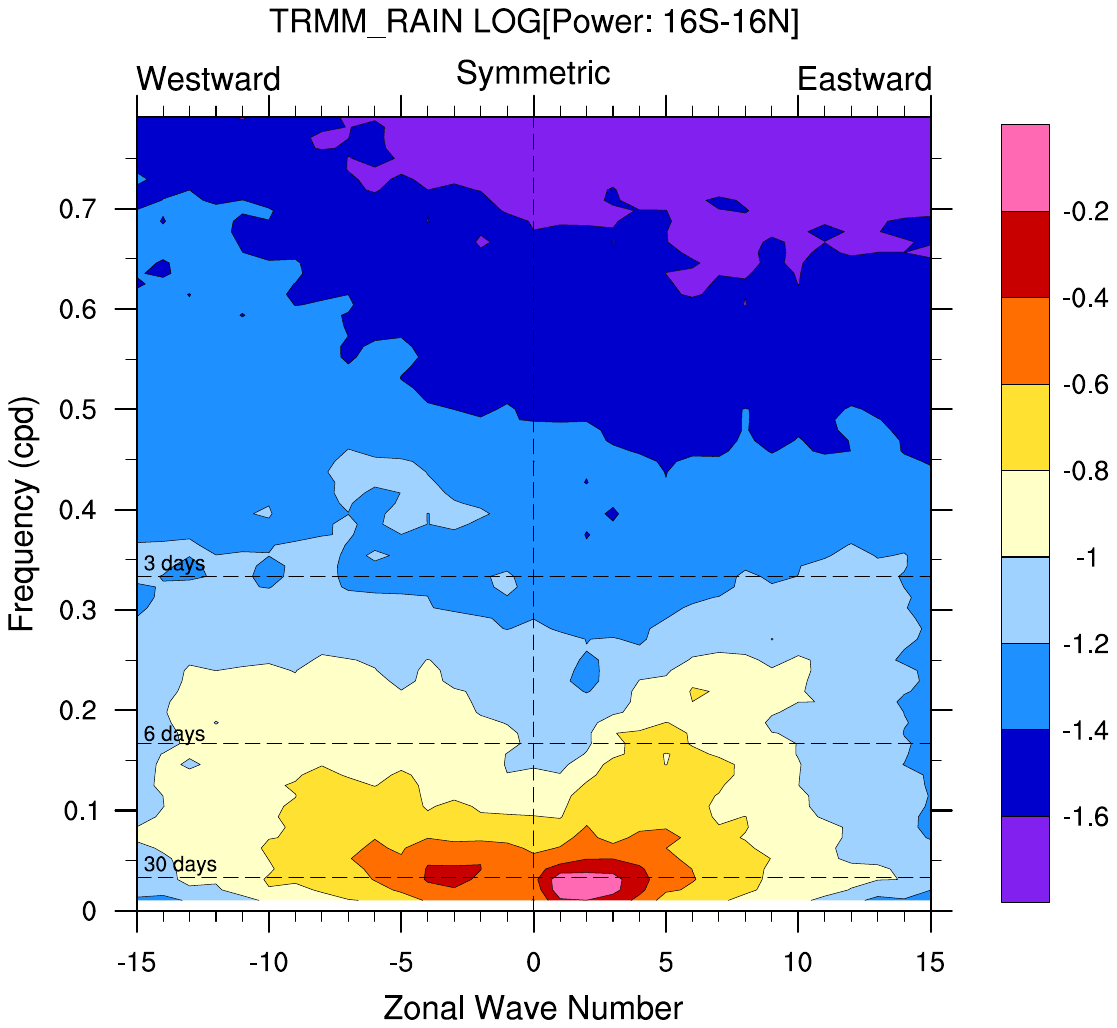}}
\caption{
Zonal wavenumber-frequency symmetric power spectra obtained from 5 years (2003 - 2007) 
of 3 hourly 16 \deg S - 16 \deg N TRMM 3B42 rainfall data. Enhanced MJO variance occurs
between zonal wavenumbers 1 and 4.
}\label{f15}
\end{figure}

Wheeler Kiladis diagrams \citep{wheeler1999} are widely used to assess the accuracy with which 
climate models simulate organized forms of tropical rainfall variance. Tropical rainfall is 
first expressed as a sum of symmetric and antisymmetric components about the equator. Each 
component is then averaged within some latitude band centered at the equator (here 16 \deg S - 
16 \deg N). Each of the two rainfall components is a function of longitude and time. This time 
series is then expressed as a power spectrum in terms of a longitudinal wavenumber $k$ and 
frequency $\nu$. In most cases, the symmetric and antisymmetric power spectra are normalized by 
a smoothed background spectrum to emphasize regions of enhanced power. Here, however, there was 
sufficient power in the simulated raw MJO spectral peak that this was unnecessary. 

Figure 14 
shows the logarithm of the power of the raw symmetric spectrum of a 300 day run of the model, 
after a 25 day spinup. For comparison, Figure 15 shows the logarithm of the power 
of the raw symmetric spectrum obtained from five years of TRMM rainfall (2003 - 2007). Both 
plots were produced using the wkSpaceTime routine of the NCAR Command Language (NCL). 
The two spectra have different scales, with the power spectrum of the model extending over a 
much larger range of amplitudes. In the TRMM spectrum, the MJO related rainfall variance is 
associated with a peak in the eastward propagating power with frequencies less than 0.05 cpd 
(i.e. periods longer than 20 days), and a range of global wave numbers from 1 to 5. The model 
spectrum exhibits a peak slightly below the dashed line showing the 30 day period, centered at 
global wave number 2. 

At any given time, the variation in 16 \deg S - 16 \deg 
N rainfall with longitude is 
not purely sinusoidal, so that higher order harmonics at all even values of zonal wave number 
are also present. The dispersion curve is clearly linear, with a slope corresponding to a phase 
speed of $6.80$ m/s (inferred from the dispersion curve intersecting the zonal wave number 15 
axis with a frequency of 0.22). Therefore, although the dominant 
frequency and wavenumber of the simulated MJO
are similar to that of the TRMM dataset, the peak in spectral space of the modeled 
MJO is much more sharply defined, and is part of a linear dispersion curve 
extending to successively weaker higher order harmonics, whose slope is roughly equal to the 
observed MJO propagation speed.


\subsection{MJO propagation speed}

The propagation speed of the MJO in the model can be inferred from the movement of the MJO 
center. In the 300 day default simulation of the model, the MJO has a speed of $6.03$ m/s to the 
east. Observed MJO's have a finite lifetime, are strongly damped \citep{lin2005}, 
and occur in a variety of SST, 
moisture, and background flow configurations. They are therefore usually characterized as having a 
range of propagation speeds of 2 - 8 m/s \citep{chen2020}.

%
%
\begin{figure}[h]
\centerline{\includegraphics[clip, trim=0.1cm 4.4cm 0.1cm 3.4cm,width=0.85\textwidth]{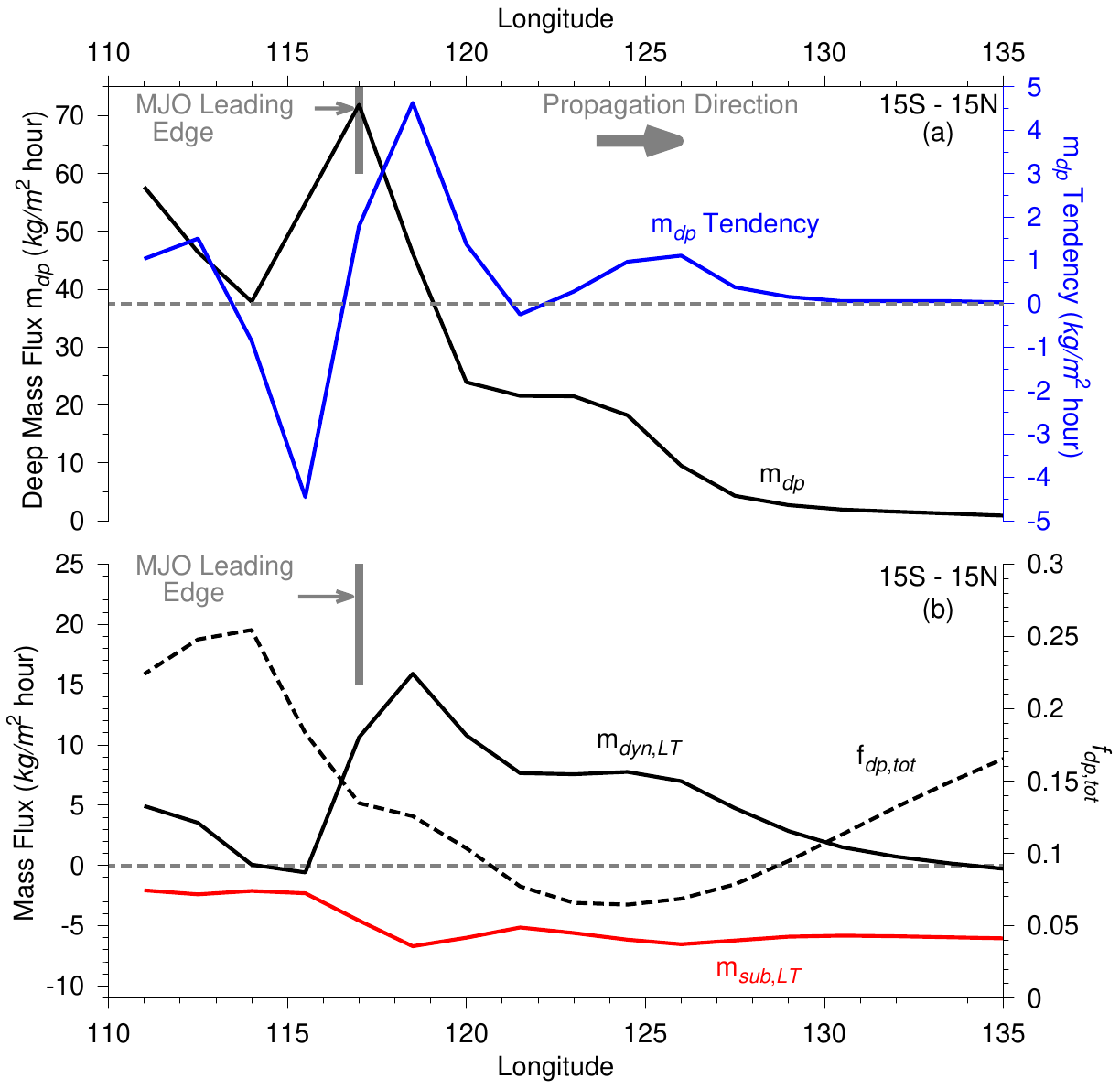}}
\caption{
(a) The top panel shows the zonal variation of the the 15 \deg S - 15 \deg N deep convective 
mass flux $m_{dp}$ (black), and the deep mass flux tendency (blue), on both sides of the 
leading eastern edge of the composite MJO. The leading edge is identified as the longitude of 
the most eastward squall line of the MJO, or most eastward $m_{dp}$ maximum. The deep mass flux 
$m_{dp}$ lags the $m_{dp}$ tendency. (b) The lower panel shows the zonal variation of the 
variables that determine the $m_{dp}$ tendency. The downward lower tropospheric subsidence mass 
flux (shown in red) is almost independent of longitude. The increase in total upward lower 
tropospheric mass flux in advance of the MJO is therefore mainly due to the dynamical component 
$m_{dyn,LT}$, shown in black. The dashed line refers to the $f_{dp,tot}$, which is the product 
of $f_{dp}(cape_{UT})$ and $f_{dp}(colrh)$. It characterizes the net effect of both upper 
tropospheric CAPE and column relative humidity on the $m_{dp}$ tendency.
}\label{f16}
\end{figure}

One unrealistic aspect of the model is that the horizontal motions generated by convective 
entrainment and detrainment occur instantaneously, as opposed to being generated by outwardly 
propagating waves with a finite phase speed. The vertical motions of the background atmosphere 
therefore respond very quickly to the mass anomalies generated by convective mass transport. 
The propagation speed of the MJO events simulated by the model is therefore constrained mainly by 
lags in the convective response to vertical motion.

The top panel of Figure 16 shows the mean longitudinal variation of the deep convective mass 
flux $m_{dp}$, and the net deep convective mass flux tendency $m_{dp,prod}$, at the eastern 
leading edge of the mean MJO pattern. The leading edge is defined as the grid cell where 
$m_{dp}$ assumes its largest value. The oscillation in the net $m_{dp}$ tendency, shown in 
blue, generates the squall line multiscale structure of the MJO. The forward advance 
of the MJO is limited by the timescale with which $m_{dp}$ can respond to the $m_{dp}$ tendency. 
In this case, the value of the deep convective mass flux at the leading edge of the MJO 
($m_{dp,LE}$) should equal the cumulative amount of deep convective mass flux production at 
the leading edge ($LE$) grid cell over the previous several days. Several days would be
sufficient because both $m_{dp}$ and $m_{dp,prod}$ relax to zero sufficiently in advance 
of the MJO envelope.
\begin{equation}
	m_{dp,LE} = \int_{days}^{previous}  m_{dp,prod} dt
\end{equation}
If we assume that the shape of the deep convective mass flux tendency is fixed relative to the 
leading edge of the MJO, and the MJO propagation speed is defined as $v_{MJO}$, we can let $dt 
= dx/v_{MJO}$, and replace the integral over time with an integral over distance along the 
equator. If the speed $v_{MJO}$ is constant, we can write
\begin{equation}
	v_{MJO} = {{\int^{east}_{of LE} m_{dp,prod} dx} \over {m_{dp,LE}}}.
\end{equation}
It is understood that the integral extends to the right of the leading edge grid cell. Using 
the values for $m_{dp,LE}$ and $m_{dp,prod}$ shown in the top panel of Figure 16 gives 
$v_{MJO} = 5.78$ m/s. This is reasonably close to the simulated MJO propagation speed 
$v = 6.03$ m/s.
Because the downdraft mass flux and the induced upward ascent from the 
downdraft circulation can both be expected to be proportional to the deep mass flux, both 
numerator and denominator in Eq. (24) should be proportional to the deep convective mass 
flux of the MJO. To first order, the MJO propagation speed should therefore not depend on 
the MJO amplitude.

In Eq. (21), the production of deep convective mass flux is proportional to the product of 
the two sigmoidal functions which capture the sensitivity of the deep mass flux tendency 
to column relative humidity and 
CAPE, i.e., $f_{dp}(colrh)$ and $f_{dp}(cape_{UT})$. The lower panel of Figure 16 shows the 
product of the two sigmoidal functions, $f_{dp,tot}$. Upper tropospheric CAPE is 
usually sufficiently large that $f_{dp}(cape_{UT})$ is reasonably close to 1. The main 
reason for the smaller values of $f_{dp,tot}$ in advance of the MJO is, therefore, that the 
column relative humidity is lower than the critical value $colrh_{half} = 0.8$ for 
$f_{dp}(colrh)$ shown in Figure 7.

The deep convective mass flux production is also proportional to the sum of the dynamical and 
subsidence vertical mass flux, $m_{dyn,LT} + m_{sub,LT}$. The mean longitudinal variation of 
these two mass fluxes in the vicinity of the leading edge of the MJO is also shown in the 
lower panel of Figure 16. The longitudinal variation in the total vertical motion is dominated 
by the dynamical component, which exhibits a peak one grid cell in advance of the leading 
edge. The propagation of the MJO can therefore be attributed to the extension of the downdraft 
congestus circulation outside the direct envelope of the MJO. In the model, one of the reasons
for the slow MJO propagation speed is because of the net subsidence between MJO events. 
MJO events must therefore continuously advance into regions of reduced column relative humidity,
within which the $f_{dp,tot}$ factor reduces the efficiency with which net upward lower tropospheric motion
can generate new deep convection. Furthermore, in order to produce a net positive deep mass flux
tendency in advance of the MJO, the downdraft circulation within the MJO envelope must first produce sufficient 
upward dynamic motion to exceed the subsidence descent.

\section{
Discussion
}

One advantage of the simple and parameterized nature of the model is that the effects of various 
processes on the behavior of the MJO events simulated by the model can be easily modified by
adjusting particular model parameters.

\subsection{
Downdraft circulation length scale
}

In the model, the deep updraft mass flux is roughly twice as large as the downdraft mass flux. 
In order for the induced lower tropospheric ascent that is part of the downdraft congestus 
circulation to locally exceed the induced descent from the deep circulation, it is necessary 
that the ascent be concentrated in a spatially smaller area than the descent. The model would 
therefore be expected to simulate MJO events only when the downdraft circulation length scale 
was significantly smaller than the deep updraft length scale (i.e., $L_{LT} < L_{UT}$). The 
upper panel of Figure 17 shows the Hovm\"{o}ller rainfall diagram that results when the 
downdraft circulation length scale is increased from the default $L_{LT} = 500$ km, to $L_{LT} 
= 700$ km. Rainfall clusters continue to appear at particular fixed latitudes. They grow 
until they reach a size comparable with an MJO, but then rapidly dissipate and do not 
propagate. If the downdraft circulation length scale $L_{LT}$ is incrementally increased from 
its default value, the latitudinal range of the MJO events tends to expand, and the MJO 
propagation speed slightly decreases. Future increases in tropical tropospheric temperatures will 
presumably give rise to an upward trend in the height of the melting level 
\citep{folkins2013}, and therefore, potential modest increases in the spatial scale of the downdraft 
congestus circulation.

%
%
\begin{figure}[h]
\centerline{\includegraphics[clip, trim=0.0cm 4.0cm 0.0cm 3.8cm,width=0.85\textwidth]{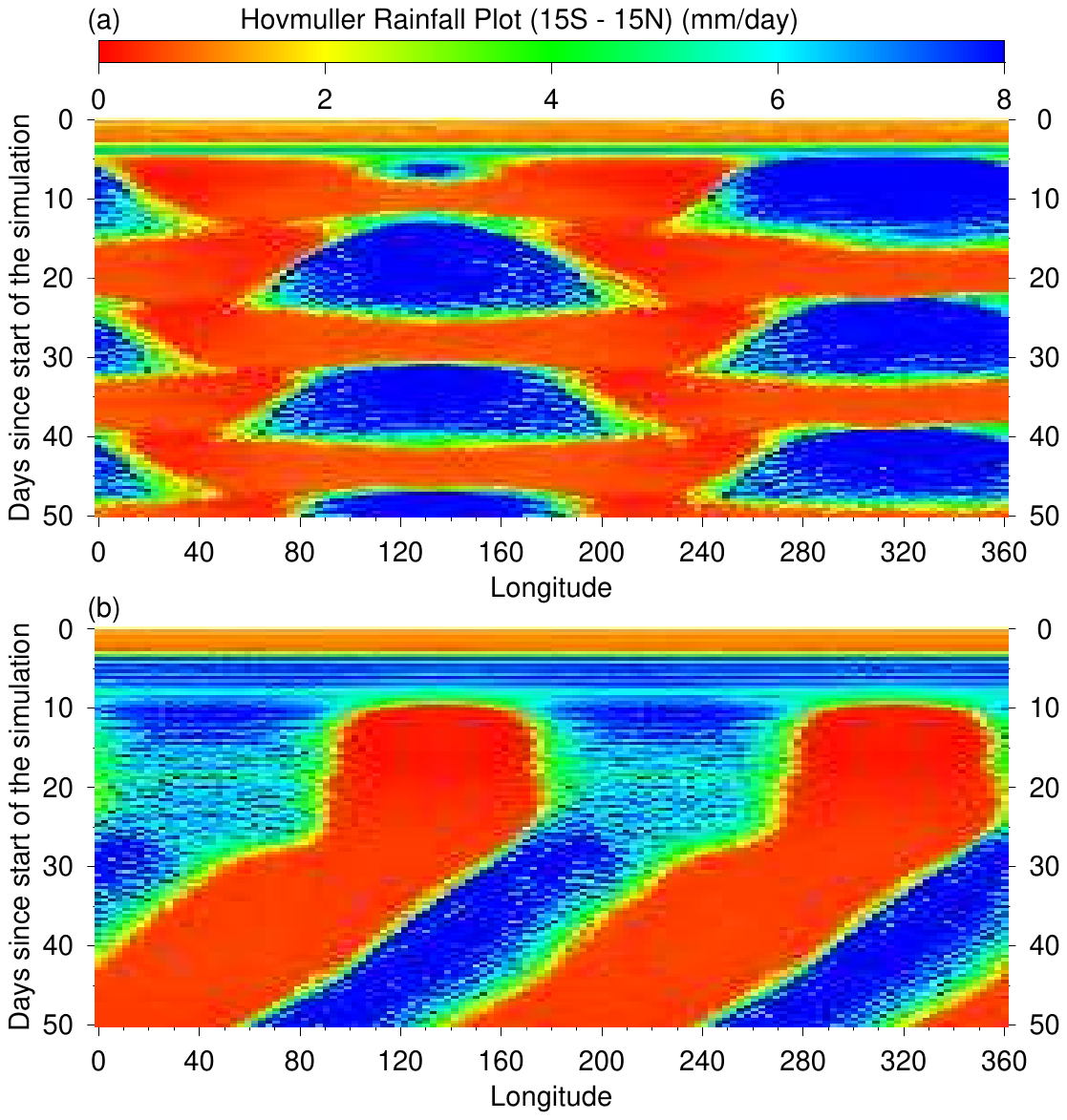}}
\caption{
(a) The top panel shows the Hovm\"{o}ller diagram of the first 50 days of a model simulation in which
the length scale of the downdraft congestus circulation $L_{LT}$ is increased from 500 km to 700 km.
(b) The lower panel shows a Hovm\"{o}ller diagram of the first 50 days of a model simulation in which
the zonal asymmetry parameter $f_{L,asym}$ is set to 1, so that the horizontal transport
caused by a local mass deficit or excess are symmetric about the eastward and westward
directions.
}\label{f17}
\end{figure}

\subsection{
Role of the Zonal Asymmetry Parameter
}

The lower panel of Figure 17 hows the Hovm\"{o}ller diagram that results when the zonal 
asymmetry parameter $f_{L,asym}$ is set to 1. In this case, the horizontal mass transport
induced by a local surplus or deficit of mass in a grid box is zonally 
symmetric. The initiation of westward propagation of the MJO events at day 25 can presumably 
be attributed to a small bias in the model which favors propagation toward the west. The 
propagation speed of the simulated MJO events, whether eastward or westward, is essentially 
independent of the value of $f_{L,asym}$. It can be shown that the effect of $f_{R,asym}$ on 
the MJO propagation direction is mainly determined by its effect on horizontal transport in 
the boundary layer. For example, increased horizontal transport in the boundary layer toward 
an MJO on the western side for $f_{R,asym} < 1$ favors increased downward dynamical
and subsidence vertical motion. This would reduce the water vapor mixing ratio of the boundary 
layer and the congestus and deep convective mass fluxes on the western side, and
favor eastern propagation.

\subsection{
Downdrafts
}

The downdraft mass flux is partially constrained by the amount of deep rainfall available to 
evaporatively cool lower tropospheric air parcels to a sufficient degree that they have a 
negative buoyancy when moved downward to the boundary layer. However, the main constraint on 
the downdraft mass flux is that it is not permitted to be larger than a prescribed fraction of 
the deep updraft mass flux. This prescribed maximum fraction increases from $f_{dn,min} = 0$ at 
low $m_{dp}$, to $f_{dn,add} = 0.8$ at large $m_{dp}$, with a transition near $m_{dp,half} = 
25$ kg/(m$^2$hour). If $f_{dn,add}$ is reduced to 0.6, the model does not exhibit MJO rainfall 
variance. If $f_{dn,add} = 0.7$, the model continues to exhibit MJO variance, but there are 
also occasional longitudinal bands of non propagating rainfall. If the downdraft fraction 
$f_{dn}$ is fixed at $0.6$ or $0.7$, rainfall clustering on the spatial scale of an MJO 
continues to occur, but there is no propagation.

\subsection{
Congestus clouds
}

The congestus mass flux is proportional to the $a_{cg}$ parameter, which determines the 
strength of the coupling with the lower tropospheric total vertical motion. The MJO variance 
in the model is considerably weakened if this parameter is reduced from the default value of 
$a_{cg} = 3$ to $a_{cg} = 1$, and is reduced to zero for $a_{cg} = 0.5$. It is not affected by 
the deactivation of the parameterization for congestus rainfall evaporation. This suggests 
that the mechanism by which congestus clouds generate MJO rainfall variance is not exclusively 
through lower tropospheric moistening.

\section{
Issues related to the coupling of convective mass fluxes to vertical motion
}

We have argued that, in constructing a convective parameterization, it is useful to pay 
particular attention to the interaction between the clear sky vertical motion and the 
convective mass fluxes. As such, the most useful diagnostic target is the observed variation in 
the vertical structure of mass divergence during the growth and decay of high rain events shown 
in Figure 12 \citep{mitovski2010}. In the model, agreement with this observational target has 
been achieved through the use of a convective parameterization in which the mid-level congestus 
mode is in phase with the net background vertical motion, while the tendency of the deep 
convective mode is in phase with the background vertical motion. This may also be a desirable 
way to couple convective mass fluxes to grid scale dynamics in models with more realistic 
dynamics. 

The vertical motion that is directly available in climate models is usually 
the pressure velocity $\omega$ that is obtained through imposing mass conservation. As such, it 
refers to the sum of all vertical mass fluxes arising from both grid scale dynamics and sub 
grid scale convective motions. However, as shown in Figure 13, even at modest rates of 
convective precipitation, the net background, or clear sky vertical motion $m_{dyn,LT} + 
m_{sub,LT}$, in the model is smaller than the updraft and downdraft convective mass fluxes, and 
can be of opposite sign to the total vertical motion. In a climate model, it may be difficult 
to determine the background clear sky vertical motion that would be most appropriate to couple 
to the convective mass fluxes. It would be inappropriate to couple the convective fluxes 
directly to the model $\omega$, however, because this would essentially couple the convective 
mass fluxes to themselves, and generate unrealistic nonlinear feedbacks.

There are other issues which arise from the way convective parameterizations are sometimes 
formulated in climate models. A convective parameterization usually consists of plumes or 
parcels that entrain air from the boundary layer, rise vertically under the influence of 
buoyancy, produce precipitation when saturated, and detrain into the free troposphere when 
they become negatively buoyant. Net convective detrainment into a grid cell increases the mass 
of that grid cell. However, above the boundary layer, the top and bottom surfaces of a grid 
cell are usually defined at fixed pressure levels. Because the convective parameterization 
must retain hydrostatic balance within a time step, the net upward convective motion of a 
column must therefore be compensated by immediate descent within the column. When convection 
is occurring in a model column, outward divergent grid scale flow usually occurs at heights 
where there is net convective detrainment, and inward convergent grid scale flow occurs where 
there is net convective entrainment. By mass continuity, convection is therefore usually 
accompanied by upward grid scale mass transport from heights where there is net entrainment, 
to heights where there is net detrainment. The descent that occurs within the convective 
parameterization is therefore largely offset by the ascent that occurs at the grid scale. 

However, the excessive and redundant vertical transport that comes from the lack of 
integration of the convective and grid scale dynamics, so that the enforcement of hydrostatic 
balance occurs individually for each process, may lead to excessive numerical diffusion in the 
vertical at higher rates of convective precipitation \citep{lawrence2008}. This issue may be 
exacerbated at higher horizontal resolutions, where the simulated rainfall variance and 
associated vertical motions are likely to be larger. However, it would be alleviated by having 
a smaller time step, so that there would be a more continuous upward flow of air within 
the convective parameterization, followed by detrainment and outward grid scale horizontal 
transport. 

In the model used here, excessive unrealistic subsidence within the grid column has 
been avoided by allowing flexibility in the upper and lower pressure boundaries of a grid 
cell. Grid cells where detrainment is occurring are permitted to absorb the additional 
detrained mass without immediate induced subsidence. It has also been alleviated by imposing
some representation of immediate outward or inward horizontal transport of some 
fraction of the local grid cell mass surplus or deficit. Although these methods may be 
inappropriate for more realistic climate models, it would likely be desirable to have some way 
of avoiding the artificially induced subsidence that occurs with some convective mass 
flux parameterizations.

\section{
Discussion and summary
}

The building block model of tropical convection \citep{mapes2006, khouider2006} has been 
implemented in a simplified model of the tropics. In this approach, the main convective heating 
profiles are considered to be those due to deep, stratiform, and congestus clouds. These basic 
heating profiles both generate and respond to the background vertical motion. Simulation of 
tropical rainfall variance requires that these basic building blocks be in correct spatial and 
temporal alignment with each other, and with the background vertical motion. The building block 
convective parameterization used here was developed using rawinsonde profiles of 
mass divergence during the growth and decay of high rain events near the 
equator as an observational target. The two most important assumptions of the convective 
parameterization are that, while the congestus mass flux is in phase with the lower tropospheric 
upward motion, the deep convective mass flux $tendency$ is in phase with the vertical motion in 
the lower troposphere.

Horizontal transport in the model is not directly generated by the pressure gradient and 
Coriolis accelerations, but is instead constrained by two main assumptions: (1) that the length 
scale for horizontal transport by the downdraft congestus circulation is roughly half that of 
the deep circulation, and (2) that the length scales of both circulations decrease with 
distance from the equator. The enforcement of these two constraints enables the simulation of an 
eastward propagating rainfall mode that exhibits many of the observed properties of the MJO. 
These include the propagation speed, internal multiscale structure, horizontal spatial scale, 
enhanced westerly inflow along the trailing edge, and enhanced cumulus congestus clouds at the 
leading edge. Higher rainfall rates within the MJO envelope are maintained mainly by increased 
variance in lower tropospheric vertical motion, and by enhanced column relative humidity. 

Forward propagation of the MJO does require, however, that some fraction of the upward motion 
generated by the downdraft circulation within the MJO propagate ahead of the MJO, and 
contribute to the net upward motion and growth of the deep convective mass flux in advance 
of the leading edge. In the model, once a lower tropospheric mass
deficit is generated within the MJO by a downdraft, 
the generation of new lower tropospheric upward motion in 
advance of the leading edge is essentially instantaneous.
The rate of forward motion of an MJO is therefore mainly 
constrained by the timescale with which the deep convective mass flux responds to this 
upward motion. In the model, the efficiency with which net upward motion in the lower
troposphere can produce deep convection is a nonlinear function of the column relative humidity.
This reduces the MJO propagation speed, because the column relative humidity between
MJO events is typically lower than 
the threshold value required to support significant deep convection.
In addition, although this is essentially enforced by fiat in the model, it is 
also probably necessary that the waves which contribute to the residual upward motion in front 
of the MJO experience some form of dissipation, since the net vertical motion generated
by freely propagating continuous waves integrates to zero over a full period.

In the model, MJO events can only move forward when the downdraft circulations that generate 
the upward motion in front of the MJO also move forward. This is in contrast with other types 
of convectively coupled waves, in which the lower tropospheric upward motion that supports the 
convection is continuously generated by the forward propagation of the wave itself as it 
interacts with the background atmospheric stability.

The main limitation of the model is that the length scales of the downdraft and deep convective 
circulations have been hardwired by parameterizations. It would be preferable that these length 
scales be shown to freely evolve from interactions between the equations of motion and the 
convective parameterization. This would then potentially allow other forms of convectively 
coupled waves to be simulated, as opposed to having almost all of the rainfall variance 
concentrated within a single mode with an unrealistically large power, as occurs in this model. 
However, the length scales used here for the deep updraft and stratiform downdraft circulations 
are significantly smaller than those obtained from the expression for the Rossby radius of a 
heat source based on its latitude and vertical depth. Therefore, depending on the horizontal and 
vertical resolution of the model, it may be necessary to introduce some form of dissipation, or 
effective reduced downdraft cooling depth, to reduce the downdraft circulation length scale to a 
value closer to that used here. 

The second limitation of the model is that the horizontal transports used to 
smooth out local grid cell mass anomalies are assumed to be occur within the one hour time 
step of the model. This is likely to be reasonably realistic for temperature, which is 
homogenized in the horizontal by gravity waves on a time scale of several hours 
\citep{sobel2001}, but it is clearly less realistic in the case of tracers such as water vapor whose 
homogenization requires physical advection.

\acknowledgments
This research was supported by the Natural
Sciences and Engineering Council of Canada.
The TRMM data were provided by the NASA/Goddard Space Flight Center's Mesoscale Atmospheric 
Processes Laboratory and PPS, which develop and compute the TMPA as a contribution to TRMM.
We thank the Atmospheric Processes And their Role in Climate (APARC) Programme
for making the HVRRD high resolution radiosonde data available for the work 
described in this article. These activities have been undertaken under the 
guidance and sponsorship of the World Climate Research Programme. Please contact the
author for the model source code, or see https://www.mathstat.dal.ca/$\sim$folkins/.

%
%

\bibliographystyle{ametsocV6}
\bibliography{references}

%
%
%

\end{document}